\documentclass{emulateapj}

\shorttitle{CARMA-COSMOS}
\shortauthors{Smol\v{c}i\'{c} et al.}

\def\f#1   {Fig.~\ref{#1}}
\def\s#1   {Sec.~\ref{#1}}
\def\tab#1   {Tab.~\ref{#1}}
\def\t#1   {Tab.~\ref{#1}}

\def\comm#1   {{\tt (COMMENT: #1) }}

\def\sqdeg            {$\Box^{\circ}$}

\def\lsun              {$\mathrm{L}_{\odot}$}
\def\lsol              {$\mathrm{L}_{\odot}$}

\def\msolyr              {$\mathrm{M}_{\odot}\, \mathrm{yr}^{-1}$}

\def\wh                {W~Hz$^{-1}$}

\def\smo               {Smol\v{c}i\'{c}}

\slugcomment{  }

\begin{document}

\title{Quest for COSMOS submillimeter galaxy counterparts using
  CARMA and VLA: Identifying three high-redshift starburst galaxies}

\author{
        V.~Smol\v{c}i\'{c}\altaffilmark{1,2,3},
        F.~Navarrete\altaffilmark{2,4}, 
        M.~Aravena\altaffilmark{5,6},
        O.~Ilbert\altaffilmark{7},
        M.~S.~Yun\altaffilmark{8},
        K.~Sheth\altaffilmark{5},
        M.~Salvato\altaffilmark{9},
        H.~J.~McCracken\altaffilmark{10},
        Diener,~C.\altaffilmark{11},
        I.~Aretxaga\altaffilmark{12},
	D.~A.~Riechers\altaffilmark{13},
	A.~Finoguenov\altaffilmark{9},
        F.~Bertoldi\altaffilmark{2},
        P.~Capak\altaffilmark{13}, 
        D.~Hughes\altaffilmark{12},
        A.~Karim\altaffilmark{14},
        E.~Schinnerer\altaffilmark{14}, 
        N.~Z.~Scoville\altaffilmark{13},
        G.~Wilson\altaffilmark{12},
        }
\altaffiltext{1}{ESO ALMA COFUND Fellow, European Southern Observatory, Karl-Schwarzschild-Strasse 2, 
85748 Garching b. Muenchen, Germany}
\altaffiltext{2}{Argelander Institut for Astronomy, Auf dem H\"{u}gel
  71, Bonn, 53121, Germany}
\altaffiltext{3}{University of Zagreb, Physics Department, Bijeni\v{c}ka cesta 32, 10002 Zagreb}
\altaffiltext{4}{Max-Planck Institut f\"{u}r Radioastronomy, Auf dem H\"{u}gel 69, 53121 Bonn, Germany}
\altaffiltext{5}{National Radio Astronomy Observatory, 520 Edgemont Road, Charlottesville, VA 22903, USA}
\altaffiltext{6}{European Southern Observatory, Alonso de C—rdoba 3107, Vitacura, Casilla 19001, Santiago 19, Chile}
\altaffiltext{7}{Laboratoire d'Astrophysique de Marseille, Universit\'e de
Provence, CNRS, BP 8, Traverse du Siphon, 13376 Marseille Cedex 12, France}
\altaffiltext{8}{Department of Astronomy, University of Massachusetts, Amherst, MA 01003, USA}
\altaffiltext{9}{Max-Planck-Institut f\"ur Extraterrestrische Physik, Giessenbachstra\ss e, 85748 Garching, Germany}
\altaffiltext{10}{Institut d'Astrophysique de Paris, UMR7095 CNRS, Universit Pierre et Marie Curie, 98 bis Boulevard Arago, 75014 Paris, France}
\altaffiltext{11}{Institute for Astronomy, ETH ZŸrich, Wolfgang-Pauli-strasse 27, 8093 ZŸrich, Switzerland}
\altaffiltext{12}{Instituto Nacional de Astrof\'{i}sica, \'Optica y Electr\'onica (INAOE), Aptdo. Postal 51 y 216, 72000 Puebla, Pue., Mexico}
\altaffiltext{13}{ California Institute of Technology, MC 249-17, 1200 East
California Boulevard, Pasadena, CA 91125 }
\altaffiltext{14}{Max Planck Institut f\"ur Astronomie, K\"onigstuhl 17,
  Heidelberg, D-69117, Germany}

\begin{abstract} We report on interferometric observations at
    1.3~mm at $2"-3"$ resolution using the Combined Array for Research
    in Millimeter-wave Astronomy (CARMA).  We identify
    multi-wavelength counterparts of three submillimeter galaxies
    (SMGs; $\mathrm{F_{1mm}}>5.5$~mJy) in the COSMOS field, initially
    detected with MAMBO and AzTEC bolometers at low, $\sim10"-30"$,
    resolution. All three sources -- AzTEC/C1, Cosbo-3 and Cosbo-8 --
    are identified to coincide with positions of 20~cm radio
    sources. Cosbo-3, however, is not associated with the most likely
    radio counterpart, closest to the MAMBO source
    position, but that further away from it. This illustrates the need for intermediate-resolution ($\sim2"$)
    mm-observations to identify the correct counterparts of single-dish
    detected SMGs. All of our three sources become prominent only at
    NIR wavelengths, and their mm-to-radio flux based redshifts
    suggest that they lie at redshifts $z\gtrsim2$. As a proof of
    concept, we show that photometric redshifts can be well determined
    for SMGs, and we find photometric-redshifts of $5.6\pm1.2$,
    $1.9^{+0.9}_{-0.5}$, and $\sim4$ for AzTEC/C1, Cosbo-3, and
    Cosbo-8, respectively. Using these we infer that these galaxies
    have radio-based star formation rates of $\gtrsim1000$~\msolyr ,
    and IR luminosities of $\sim10^{13}$~\lsol \ consistent with
    properties of high-redshift SMGs.  In summary, our sources reflect
    a variety of SMG properties in terms of redshift and clustering,
    consistent with the framework that SMGs are progenitors of
    $z\sim2$ and today's passive galaxies.
\end{abstract}


\section{Introduction}
\label{sec:intro}

Submillimeter galaxies (SMGs) are ultra-luminous, dusty star-bursting
systems with extreme star formation rates in the range of
$\sim100-1000$~M$_\odot$~yr$^{-1}$ (e.g.\ Blain 2002). It
has been shown that 
the bulk of this population is between z$\sim$2
and 3 (e.g.\ Chapman et al.\ 2005). But recently a possible
high-redshift tail of SMGs has started to emerge (e.g.\ Younger et
al.\ 2007, 2009; Valiante et al.\ 2007).
To date, only about ten $z>4$ SMGs have been confirmed \citep{daddi09a,daddi09b, capak08,schinnerer08,riechers10,capak10, smo11, coppin09,coppin10,knudsen10,cox11,combes12}. Their number density is still consistent
  with that expected in cosmological models \citep{baugh05, coppin09,
    smo11}. Note however that these $z>4$ studies are not complete,
  and may even point to the existence of a new or different SMG
  population \citep{wall08}. The intense starburst that creates the
submillimeter bright emission is likely to occur when the bulk of the
stellar mass is being assembled in these galaxies; SMGs are generally
believed to be the progenitors of today's massive red-and-dead
elliptical galaxies which formed in an intense burst at high redshift
\citep[e.g.][]{cimatti08}.  It is therefore critical to study in
detail these cosmologically important objects.

SMGs are generally detected in mm and sub-mm surveys with single-dish
telescopes that have large beams ($>10"$).  The next step is then to
pinpoint the precise locations of these objects and to match them with
their multi-wavelength counterparts and obtain a redshift.
Finding the real counterpart for an SMG is not trivial because the
spatial density of optical/IR galaxies in deep fields is high and
usually there are multiple galaxies within one single dish mm/sub-mm
beam. Deep radio, mid-IR, optical and UV (and hard X-rays for AGN)
data of higher resolution have been used to identify the right
counterpart by tracing the bright star formation or AGN activity
(e.g.\ Ivison et al.\ 2007).  However, depending on galaxy properties
and redshift, these different tracers are likely to introduce
identification-biases, i.e.\ provide true identifications for only a
fraction of the sample (and likely at low redshifts). In essence, the
most efficient and least biased way to associate counterparts is
through high-resolution mm observations. This has to date been a
  time-consuming process that resulted in a total of $\sim50$ SMGs
  detected via mm-interferometry (Downes et al.\ 1999, Frayer et al.\
  2000; Dannerbauer et al.\ 2002; Downes \& Solomon 2003; Genzel et
  al.\ 2003; Kneib et al.\ 2005; Greve et al.\ 2005; Tacconi et al.\
  2006; Sheth et al.\ 2004, 2012, Iono et al.\ 2006; Younger et al.\
  2007, 2009; Aravena et al.\ 2010; Ikarashi et al.\
2011; Tamura et al.\ 2010; Katsukade et al.\ 2010; Wang et al.\ 2011;
Chen et al.\ 2011; Neri et al.\ 2003; Chapman et al.\ 2008; Hatsukade et al.\ 2010). To date the largest
  comprehensive sample of SMGs detected at high-resolution via
  mm-interferometry is that in the COSMOS field (Younger et al.\ 2007,
  2009, Aravena et al.\ 2010) and consists of $\sim20$ sources in total. 
Here we present 1.3~mm imaging at $\sim2-3"$ resolution with the CARMA
interferometer of 3 further SMGs ($\mathrm{F_{1mm}}>5.5$~mJy) in the
COSMOS field originally detected in the MAMBO- and AzTEC-COSMOS
surveys (Bertoldi et al.\ 2007; Aretxaga et al.\ 2011). The
counterpart association based on previous data was highly ambiguous
due to multiple or faint potential radio counterparts lacking
optical/NIR detections.  We adopt $H_0=70$, $\Omega_M=0.3$,
$\Omega_\Lambda=0.7$.

\section{Data}
\label{sec:data}

\subsection{COSMOS survey}

The COSMOS project is a panchromatic (X-ray to radio) survey of an
equatorial 2\sqdeg\ field. The field has been observed with the major
space- (Chandra -- \citealt{elvis10}; GALEX -- \citealt{zamojski07};
HST -- \citealt{scoville07, koekemoer07, leauthaud07}; Spitzer --
\citealt{sanders07, ilbert10, lefloch09, frayer09}) and ground-based telescopes (Subaru, CFHT, UKIRT,
NOAO -- \citealt{capak07,taniguchi07,mccracken10}; VLA \citealt[\smo\ et
al.,~in~prep.]{schinnerer07,schinnerer10}) in more than 30 bands.
Here we additionally use deep UltraVista observations in Y, J,
  H, Ks bands taken between Dec/2009 and Apr/2010 (McCracken et al., in
  prep.). Fractions of the field have been surveyed at (sub-)mm
wavelengths (Bertoldi et al.\ 2007; Scott et al.\ 2008; Aretxaga et
al.\ 2011).
The three SMGs targeted here were initially detected by MAMBO
(Bertoldi et al.\ 2007) and AzTEC/ASTE (Aretxaga et al.\ 2011) surveys
of the COSMOS field. The
deboosted $\sim1$~mm fluxes are $13.0^{+1.1}_{-1.0}$~mJy (Aretxaga et al.\
2011), $7.45\pm1.1$~mJy, and
$5.45\pm1.0$~mJy (Bertoldi et al.\ 2007) for AzTEC/C1, Cosbo-3, Cosbo-8,
respectively (see also \t{tab:det} ).

\begin{table*}
\begin{center}
\caption{Summary of observations with the CARMA interferometer at
  230~GHz in D-configuration.}
\label{tab:obs}
\vskip 10pt
\begin{tabular}{ccccccc}
\hline
    Source       &   Pointing position  &  Date  &
    On-source & beam & rms \\
          & [J2000]  &   
    &   time [hr] & [arcsec] & [mJy/beam]  \\
\hline 
      AzTEC/C1                    &   10 01 41.68 \, +02 27 11.80 &
      Mar/2009 &   1.5 & $4.2''\times3.1''$ & 2.3 \\
      Cosbo-3                      &   10 00 57.20 \, +02 20 13.00 &
      Feb/2009 &   4.0 & $2.7''\times1.9''$ & 0.7 \\
      Cosbo-8                      &   10 00 00.00 \, +02 06 34.00 &
      Mar/2009 &   5.4 & $2.6''\times2.4''$ &
      1.5 \\
\hline
\end{tabular}
\end{center}
\end{table*}

\begin{table*}
{\footnotesize
\begin{center}
\caption{CARMA detections and VLA counterparts}
\label{tab:det}
\vskip 10pt
\begin{tabular}{ccccccccl}
\hline
    Source        & CARMA position &  $F_\mathrm{1.3mm}$
    &$F_\mathrm{1.2mm}$ & $F_\mathrm{1.1mm}$ & VLA 
    & $F_\mathrm{1.4GHz}$ & photo-z & photo-z\\
    name & [J2000] & [mJy] &  [mJy] &  [mJy] & distance &
    [$\mu$Jy] & mm/radio& UV-MIR\\
\hline 
      AzTEC/C1$^\mathrm{a}$                    &   10 01 41.75  \, +02 27 13.06 &
      $7.4\pm2.3^*$ & -- & $13^{+1.1}_{-1.0}$ & 0.28'' & $44\pm10^{+}$ & $4.3^{+0.7}_{-1.4}$ & $5.6\pm1.2$
      \\
      Cosbo-3$^\mathrm{b}$                      &   10 00 56.95  \, +02 20 17.79 &
      $5.4\pm0.7^{**}$& $7.45\pm1.1$ &$9.6^{+1.1}_{-1.0}$ & 0.25'' & $78\pm13^{+}$ & $3.2^{+0.6}_{-1.0}$ & $1.9^{+0.9}_{-0.5}$
       \\
      Cosbo-8$^\mathrm{b}$                      &    09 59 59.92  \, +02 06 33.41 &
      $4.8\pm1.5^*$& $5.45\pm1.0$ & $3.7^{+1.1}_{-1.2}$ & 0.35'' &
      $104\pm13^{+}$  & $1.9^{+0.5}_{-0.7}$ & $\sim4$
      \\
\hline
\end{tabular}
\end{center}
$^*$ Peak intensity\\
$^{**}$ Integral intensity given; the peak intensity is 2.8~mJy \\
$^{+}$ Adopted from the VLA-COSMOS catalogs (Schinnerer et al.\ 2007,
2010) \\
$^\mathrm{a}$ Aretxaga et al.\ (2011)\\
$^\mathrm{b}$ Bertoldi et al.\ (2007)
}
\end{table*}

\subsection{CARMA observations, data reduction and source detection}

We observed the three sources at 1.3\,mm using the CARMA
interferometer in a compact -- D array -- configuration. The targets
-- AzTEC/C1, Cosbo-3, and Cosbo-8 -- were observed with 15\,antennas
(corresponding to 105\,baselines) in Feb./Mar., 2009 for a total
on-source time of 1.5, 4.0, and 5.4 hours, respectively. Weather
conditions varied between acceptable and very good for 1\,mm
observations. The nearby quasar 1058+015 was observed every
15\,minutes for secondary amplitude and phase calibration. The strong
calibrator sources 3C\,84, 3C\,273, and 0854+201 were observed at
least once per track for bandpass and flux calibration. Radio pointing
was performed at least every 2.5\,hr on nearby sources. The resulting
total flux calibration is estimated to be accurate within 15--20\%.

The upper (lower) sidebands of the 1\,mm receivers were centered at
230~(225)~GHz. Each sideband was observed with 45 channels each
31.25\,MHz wide, for a total bandwidth of 2.8\,GHz
(2$\times$1406.25\,MHz).  For data reduction and analysis, the MIRIAD
package was used. The final data cube obtained after flagging (and
combination of the datasets from all runs) was collapsed along the
frequency axis to obtain 1.3\,mm continuum images. The $u-v$ data were
imaged with natural baseline weighting, leading to synthesized clean
beam sizes (rms values) of $4.2''\times3.1''$ (2.3~mJy/beam),
$2.7''\times1.9''$ (0.7~mJy/beam), and $2.6''\times2.4''$
(1.5~mJy/beam) for AzTEC/C1, Cosbo-3, and Cosbo-8, respectively.  The
observations are summarized in \t{tab:obs} .

The 1.3~mm CARMA stamps are shown in \f{fig:stamps} . All three
sources are detected at a $\sim3-4\sigma$ level. We stress that the
positions of our 1.3~mm sources perfectly coincide (within
$\lesssim0.3"$) with significant 20~cm (1.4~GHz) radio detections
drawn from the VLA-COSMOS survey \citep[][see \t{tab:det}
]{schinnerer07,schinnerer10}. As the chance probability of finding a
radio source within the CARMA beam (given the radio source number
density) is of the order of only $\sim10^{-4 }$, this significantly
boosts the validity of our mm detections.  We have extracted the
1.3~mm (230~GHz) fluxes using the AIPS tasks MAXFIT (that identifies
the position with maximum value in a selected pixel array) and JMFIT
(that fits a 2D Gaussian to selected pixel arrays). The flux densities
and the corresponding errors are summarized in \t{tab:det} . The
fluxes are in relatively good agreement (i.e.\ within $1\sigma$) with
those inferred from the MAMBO data, and show a stronger deviation from
the AzTEC 1.1~mm data.  This is likely due to a steep spectral index in
the rest-frame sub-mm band which translates into a rapid change in flux
even within the 1~mm window. Assuming $\beta=1.0$ and $2.0$ we expect
a factor between 1.6 and 1.9 discrepancy between the observed
1.3~mm and 1.1~mm flux densities.

\begin{figure*}
\includegraphics[scale=0.65]{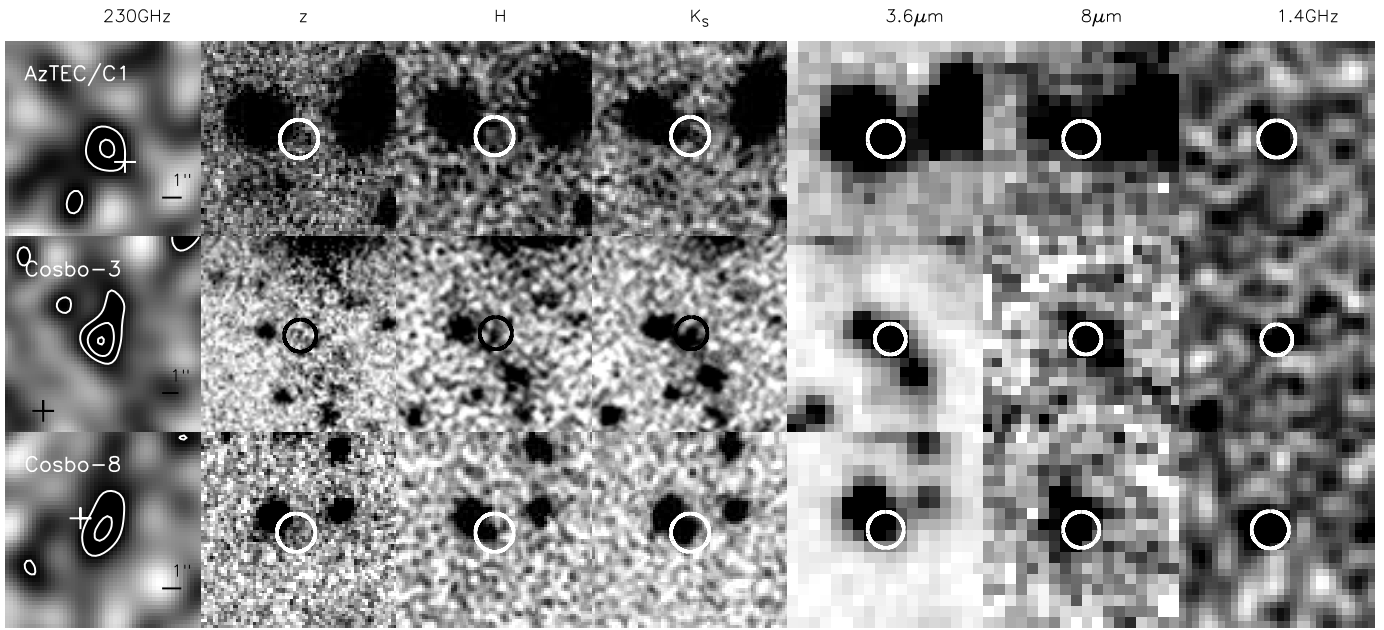}
\caption{ NIR to radio stamps for AzTEC-C1, Cosbo-3, and Cosbo-8 with
  1.3~mm (CARMA) contours overlayed. The contour levels are at
  $2\sigma$, $3\sigma$, and $4\sigma$ ($1\sigma=2.3, 0.7,
  1.5$~mJy/beam for AzTEC-C1, Cosbo-3, and Cosbo-8, respectively). The
  cross indicates the pointing center at 230~GHz. The circle ($2"$ in diameter) indicates the CARMA source position. }
	\label{fig:stamps}
\end{figure*}

\section{Analysis and results}
\label{sec:results}

\subsection{Multi-wavelength counterparts and their photometry}

Images of the three SMGs at various wavelengths, with 1.3~mm CARMA
contours overlaid, are shown in \f{fig:stamps} . All three
  sources are coincident (within $\lesssim0.3"$) with 20~cm radio
  detections as summarized in \t{tab:det} . Only Cosbo-8 is
  detected in X-ray emission (within
  the Chandra-COSMOS survey; Elvis et al.\ 2010). The multi-wavelength
  counterparts of our sources become prominent only at NIR
  wavelengths, and are blended by nearby sources. In order to extract
  the most accurate photometry for these sources we have carefully
  deblended the SMG emission (up to the Spitzer/IRAC 8~$\mu$m band), and
  extracted its flux as described in the following.
 
\subsubsection{Calibration of photometry using the COSMOS photometric-catalog}
\label{sec:phot}

\begin{table}
\begin{center}
\caption{Photometry calibration}
\label{tab:apert}
\vskip 10pt
\begin{tabular}{cccc}
\hline
    band       &   Aperture  & mag & mag \\
    &   diameter [$"$] &  offset & error\\
\hline 
g & 3 & 0.1123 & 0.034\\
r & 3 & 0.2514 & 0.016\\
i & 3 & 0.2619 & 0.034\\
z & 3 & 0.2203$^a$ & 0.039\\
J & 3 & 0.0828 & 0.063\\
H & 2.3 & 0.0644 & 0.054\\
K$_s$ & 2 & -0.0390 & 0.080\\
Y-UltraVista & 2 & 0.0000 & 0.040\\
J-UltraVista & 2 & -0.0225 & 0.045\\
H-UltraVista & 2 & -0.1111 & 0.035\\
K$_s$-UltraVista & 2 & -0.1225 & 0.030\\
3.6~$\mu$m$^*$ & 3.8 & 0.0729$^{b}$ & 0.048\\
4.5~$\mu$m$^*$ & 3.8 & 0.0724$^{c}$ & 0.038\\
5.8~$\mu$m$^*$ & 3.8 & 0.0807 & 0.063\\
8.0~$\mu$m$^*$ & 3.8 & 0.0823 & 0.148\\
IA427 & 2.5 & 0.0828 & 0.052\\
IA464 & 2.5 & -0.067 & 0.049\\
IA484 & 2.5 & 0.1418 & 0.067\\
IA505 & 2.5 & 0.1105 & 0.060\\
IA527 & 2.5 & 0.1210 & 0.072\\
IA574 & 2.5 & -0.0172 & 0.042\\
IA624 & 2.5 & 0.1786 & 0.051\\
IA679 & 2.5 & 0.0301 & 0.055\\
IA709 & 2.5 & 0.0173 & 0.050\\
IA738 & 2.5 & 0.1119 & 0.055\\
IA767 & 2.5 & 0.0263 & 0.087\\
IA827 & 2.5 & -0.0875 & 0.036\\
\hline
\end{tabular}
\end{center}
$^*$ total (i.e.\ aperture corrected) magnitude \\
$^{a}$ for the computation of the z-band magnitude we adopt the
relation presented in McCracken et al.\ (2010), and scale it
additionally with the offset magnitude given here\\
$^{b}$ in addition, for $m_\mathrm{3.6\mu m}>21.5$ we require $m_\mathrm{ap}=1.215\,m_\mathrm{ap}-5.356$\\
$^{c}$ in addition, for $m_\mathrm{3.6\mu m}>21.5$ we require $m_\mathrm{ap}=1.310\,m_\mathrm{ap}-6.770$
\end{table}

In this section we describe our photometric extraction procedure that will be applied in the next section to the CARMA-COSMOS UV-MIR counterparts. 
The photometry of sources in the COSMOS photometric catalog is
extracted using aperture techniques, which we also adopt here.  To
validate our photometric extraction procedure, and to estimate its
uncertainty we have drawn $\sim100$ random sources from the i-band
selected catalog \citep{capak07} that also have IR detections, and are
outside masked regions in the field. Using images at their original
resolution (not convolved to a common FWHM; see \citealt{capak07} for
details), we adopt aperture sizes for individual bands
as summarized in \t{tab:apert} . These aperture sizes were chosen to
achieve the best agreement with the photometry reported in the COSMOS
photometric catalog. We also corrected our aperture magnitudes for
slight systematic offsets (see \t{tab:apert} ) in order to put them on
the COSMOS photometry scale. Thus, our final magnitudes are computed
as $m^\mathrm{final}_\mathrm{apert}=m_\mathrm{apert}+m_\mathrm{offset}$. Note also,
that for the IRAC 3.6 and 4.5~$\mu$m bands we fit the faint magnitude
slope separately (as reported in \t{tab:apert} ), and for the Subaru
$z^+$ band we adopt the relation given by \citet{mccracken10}.

We estimate the average error of our magnitudes
by statistical propagation of the magnitude errors reported in the
catalog and the spread of the (catalog - aperture) magnitude
difference (in cases where the average error of the catalog magnitudes
is larger than the spread we adopt the latter as the error of our
magnitudes). In summary, our aperture photometry matches that in the
COSMOS photometric catalog very well (mean offsets are zero), and the
average error of such extracted magnitudes is estimated to be
$\sim0.05$ (see \t{tab:apert} \ for exact values for each band).

\subsubsection{Deblending}
\label{sec:debl}

\begin{figure*}
\includegraphics[bb= 0 200 432 302, scale=0.6] {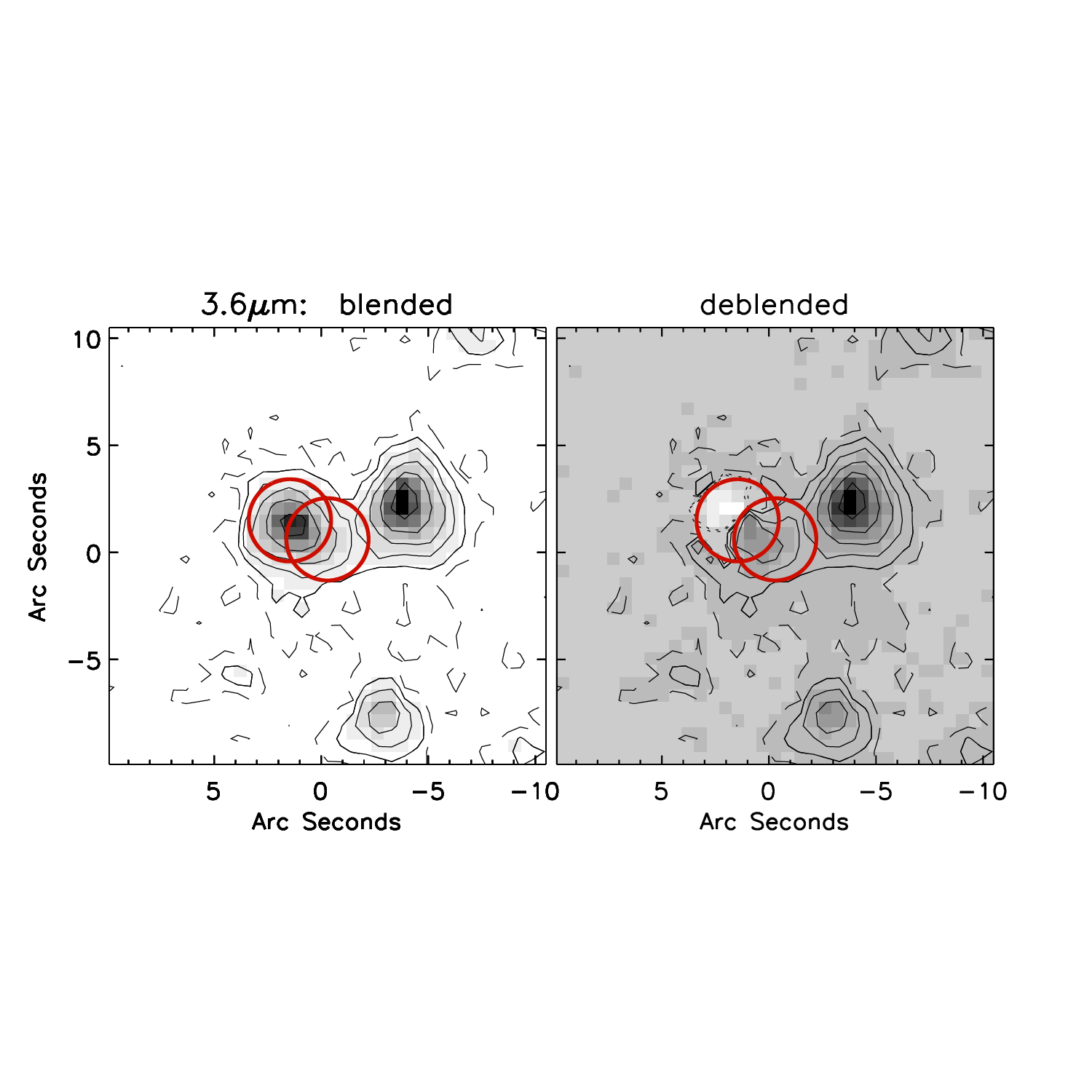}
\includegraphics[bb= 0 200 432 432, scale=0.6] {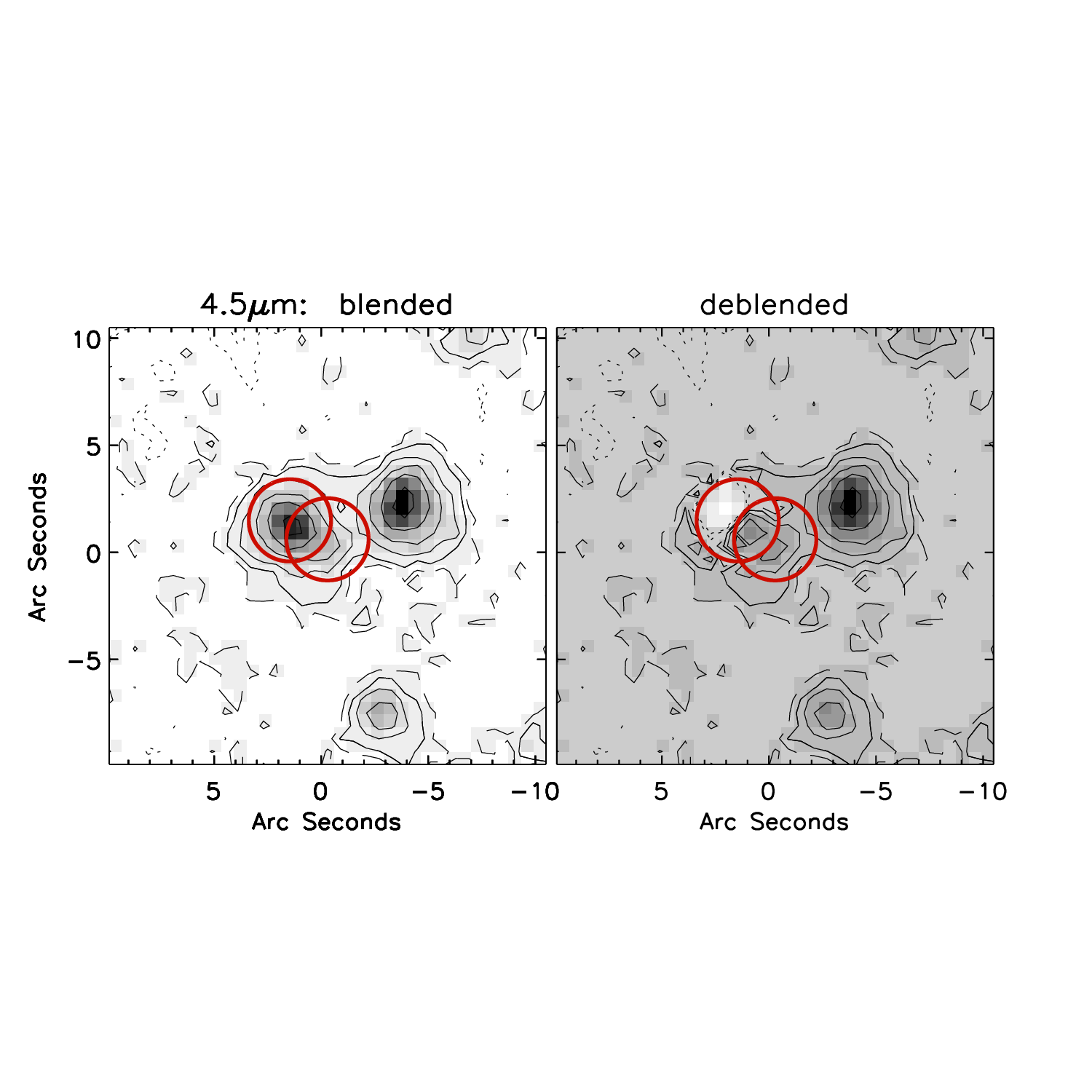}\\
\includegraphics[bb= 0 150 432 432, scale=0.6] {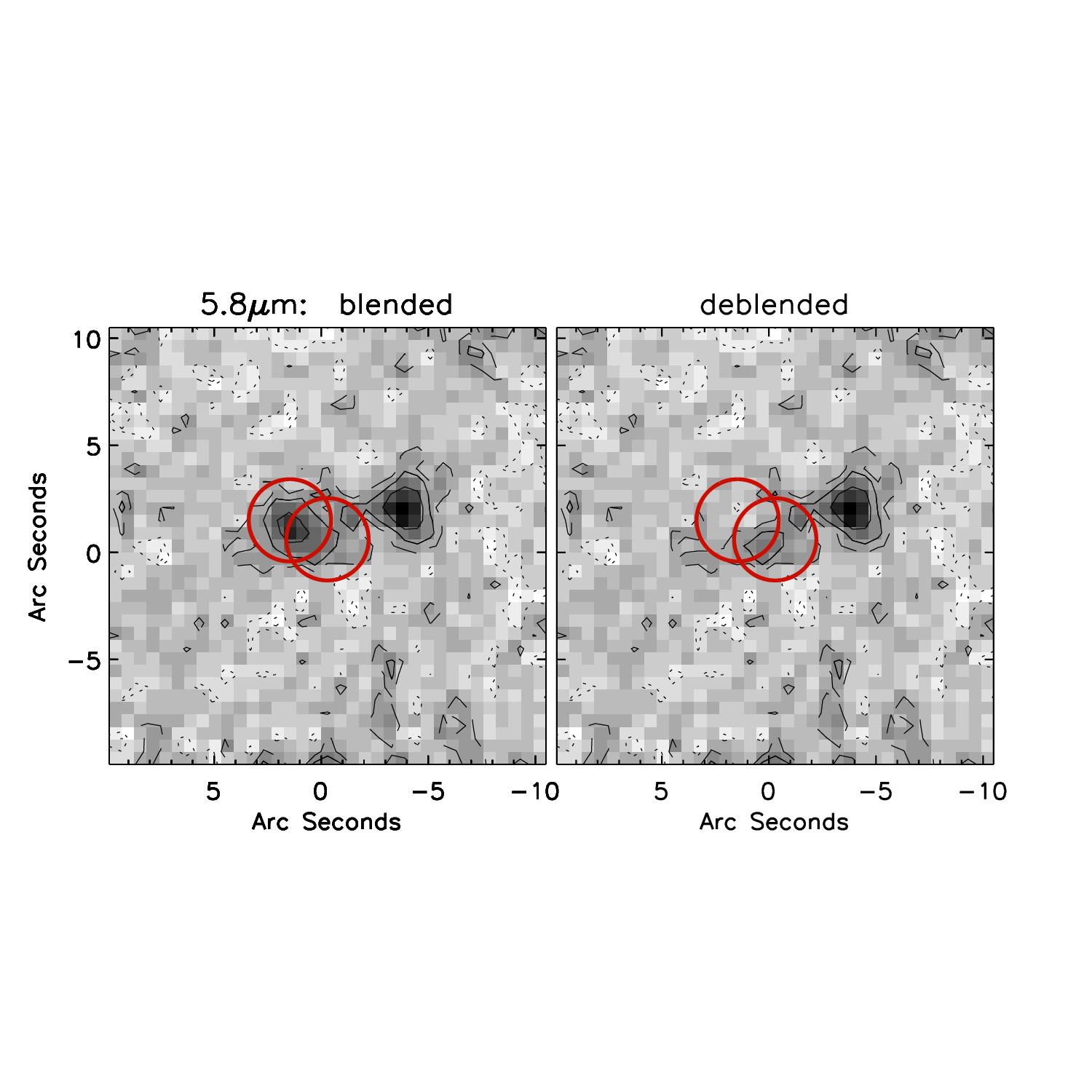}
\includegraphics[bb= 0 150 432 432, scale=0.6] {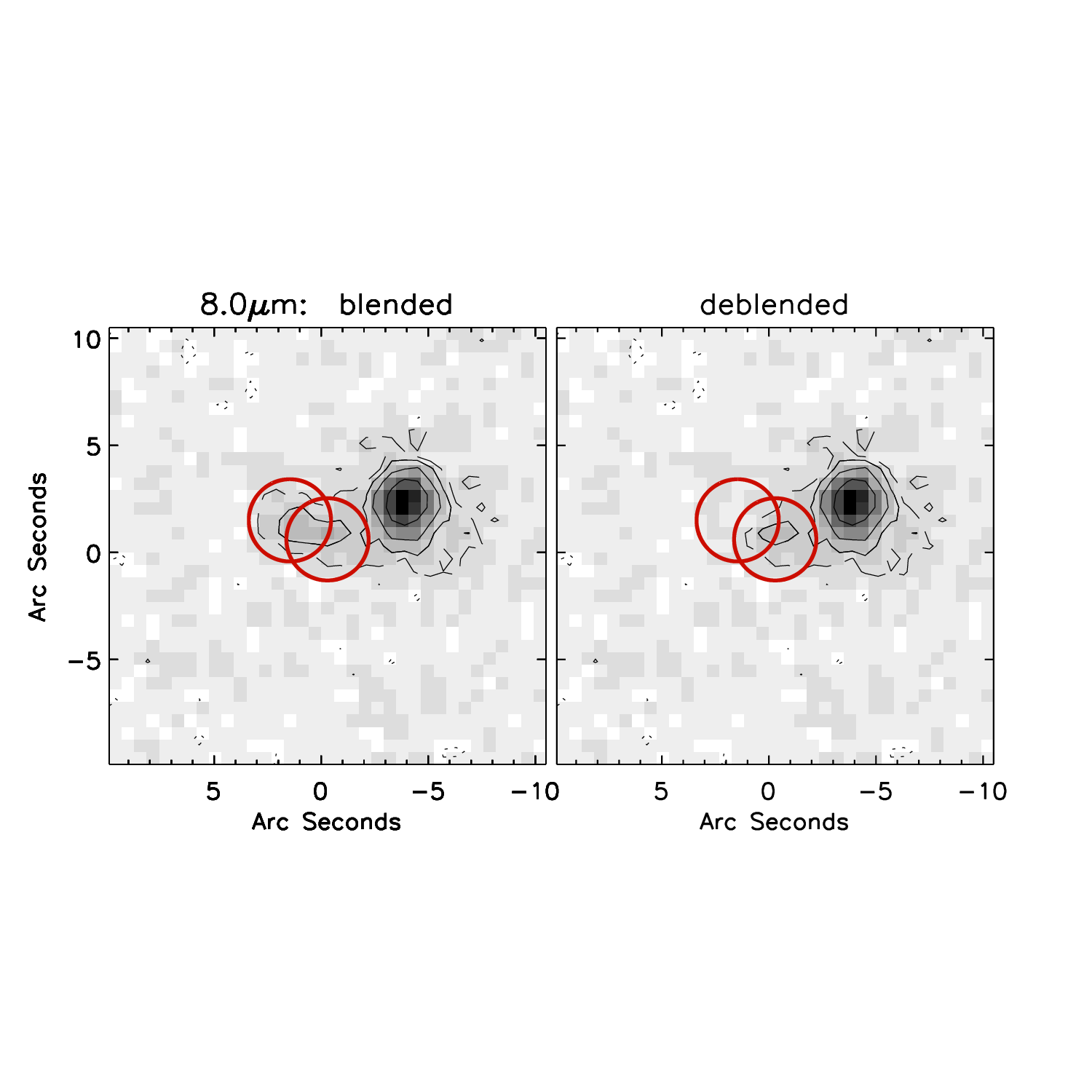}\\
                           \\ \\ \\ 
\caption{AzTEC/C1: IRAC 3.6~$\mu$m (top left), 4.5~$\mu$m (top right), 5.8~$\mu$m (bottom 
  left), 8.0~$\mu$m (bottom right) blended (left panel column) and
  de-blended (right column) stamps. The contour levels are at $-4\sigma,\, -2\sigma$
  (dotted lines), $2\sigma,\,4\sigma$ (dashed lines) and $2^i\sigma, i=3,4,5,6...$ (full lines),
  where the rms ($\sigma$) has been derived locally. The CARMA position is marked by
  the thick circle in the center of the stamp (its size matches the photometry
  aperture for the given band; see \t{tab:apert} ). The blending source subtracted is outlined by the circle
  to the NE. In the 3.6, and 4.5 $\mu$m bands the deblending was performed by mirroring
  the NE source across its SE-NW diagonal within a $2''$ aperture
  (which artificially induces the negative area within the aperture in
  the deblended images). To deblend the 5.8, and 8.0~$\mu$m bands a point source was
  subtracted. }
	\label{fig:deblendc1}
\end{figure*}

\begin{figure*}[t!]
\includegraphics[bb =0 200 432 300, scale=0.6]{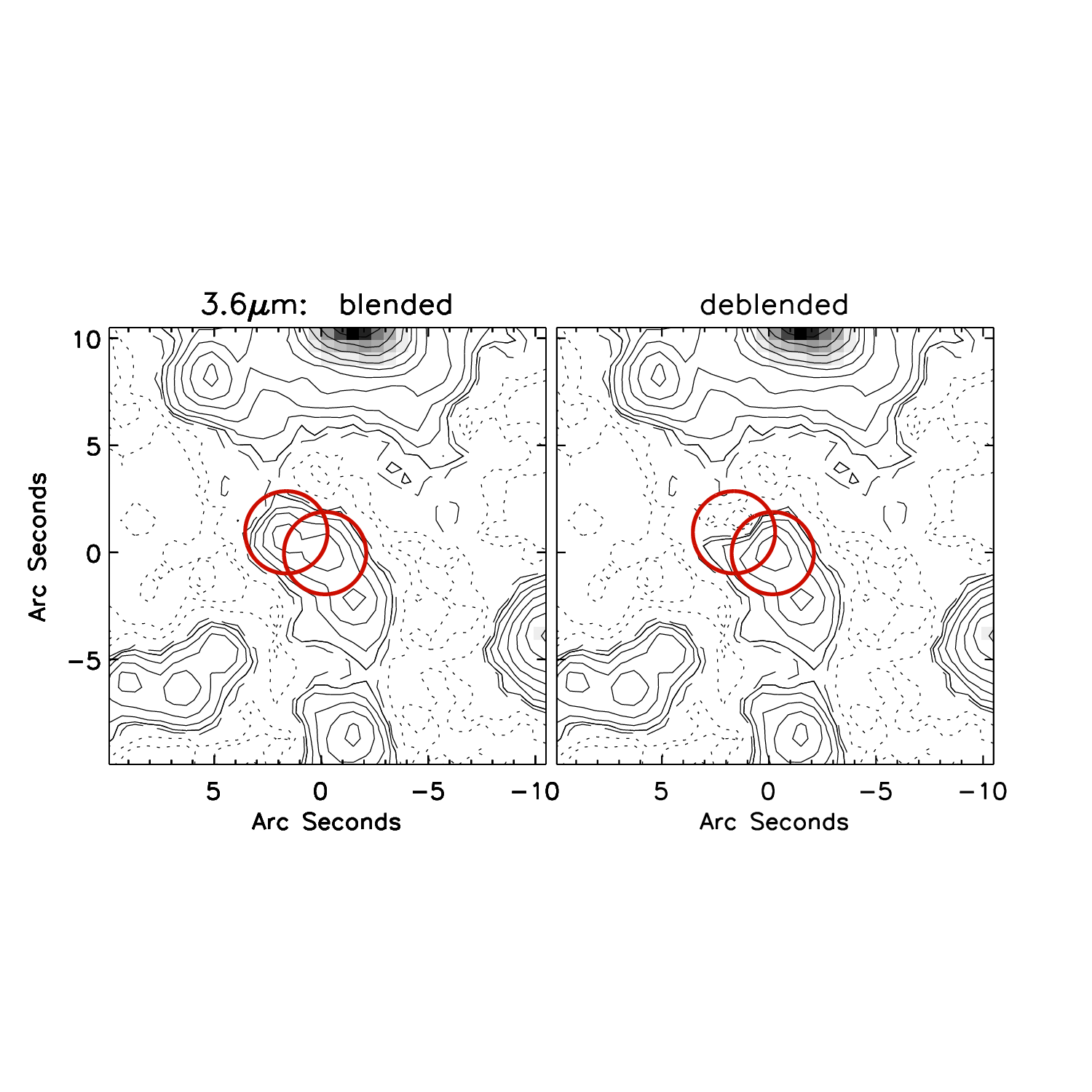}
\includegraphics[bb =0 200 432 300, scale=0.6] {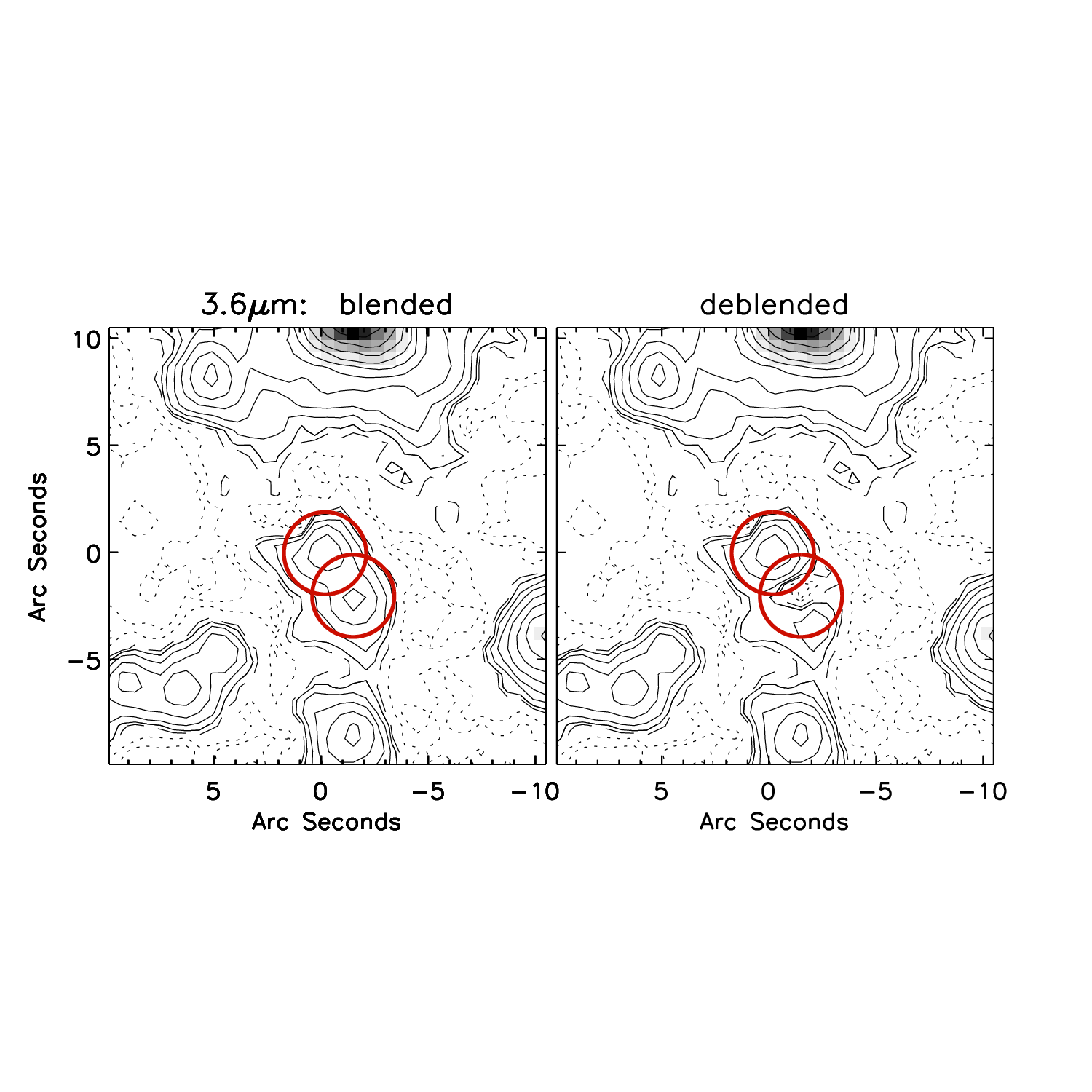}
\includegraphics[bb =0 200 432 432, scale=0.6] {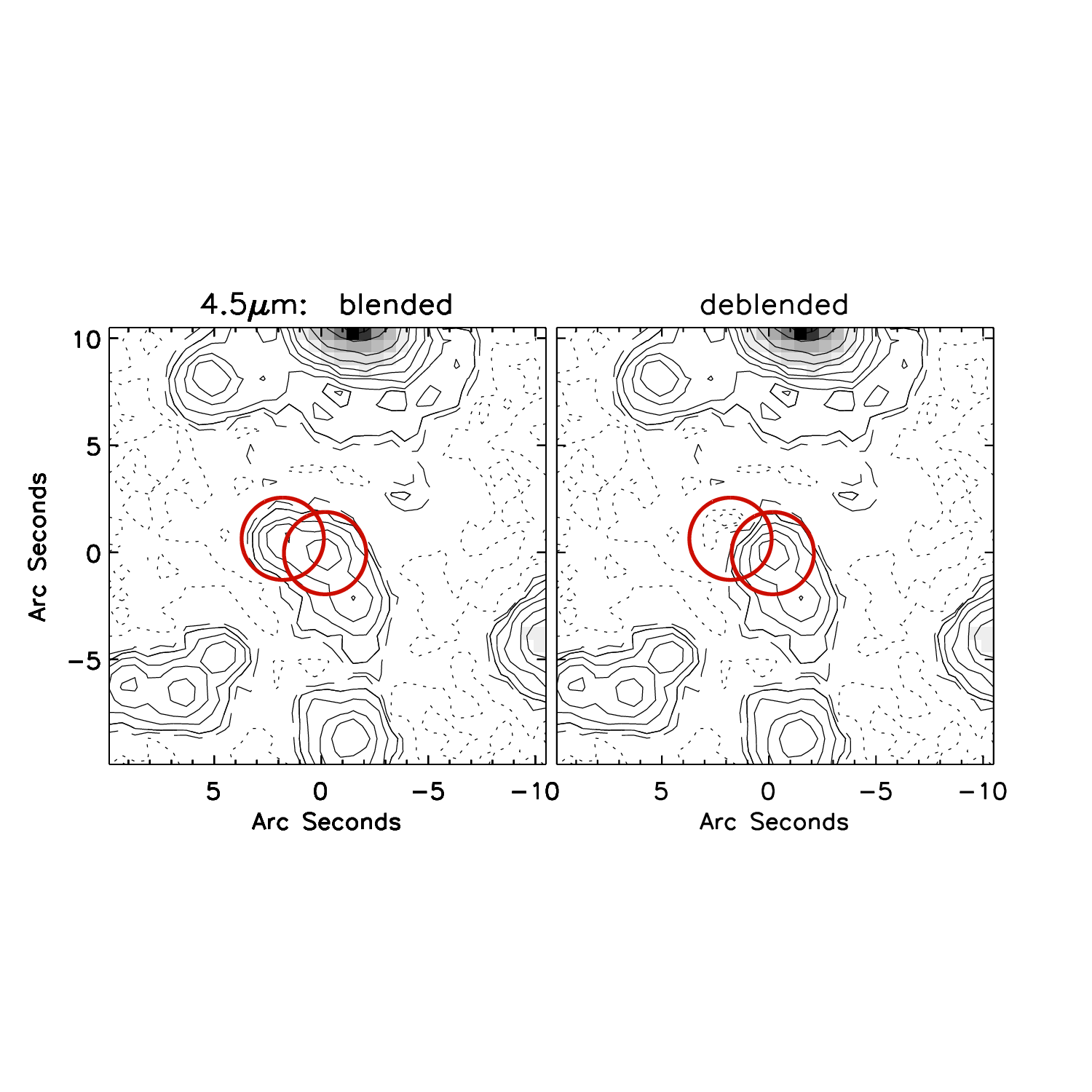}
\includegraphics[bb =0 200 432 432, scale=0.6] {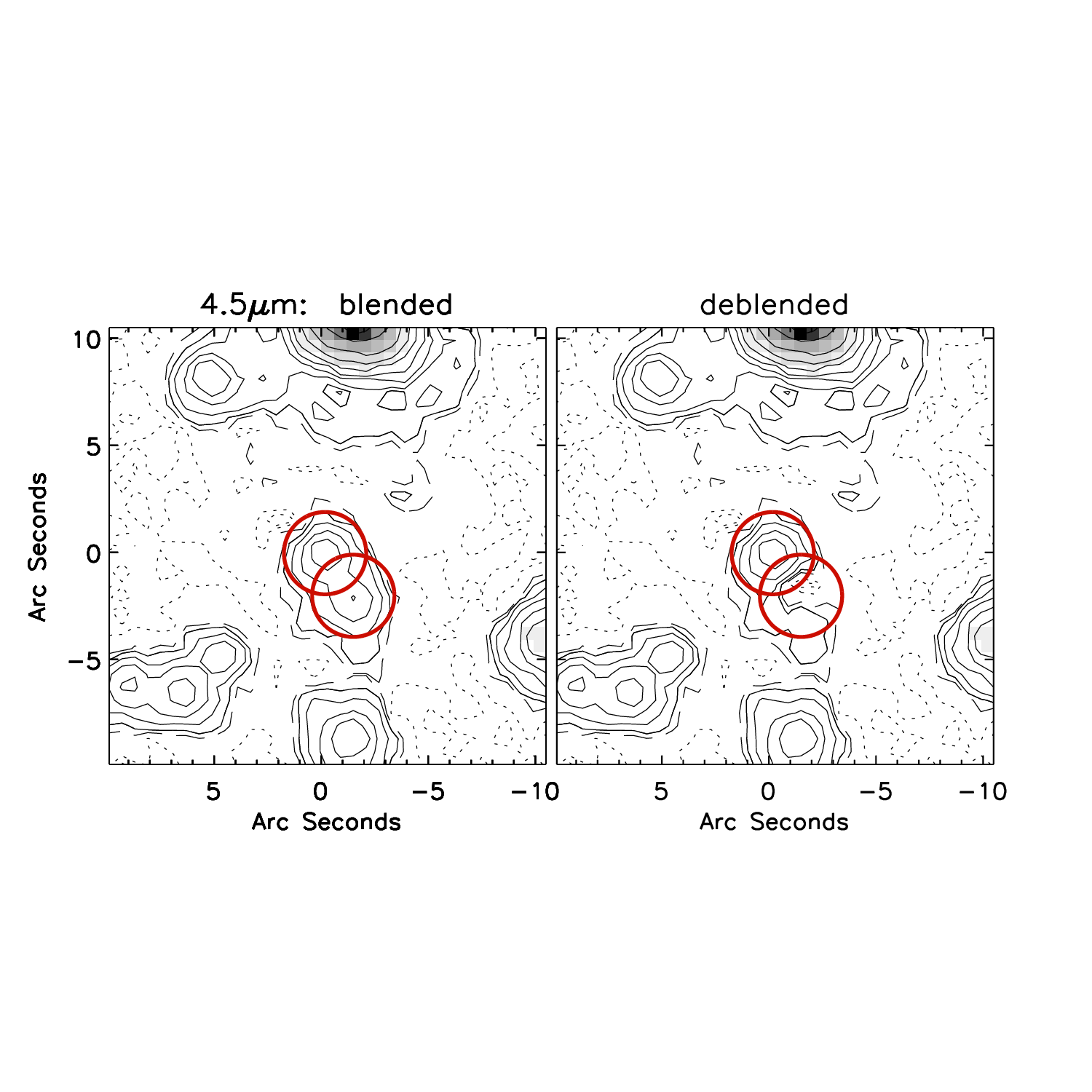}
\includegraphics[bb =0 200 432 432, scale=0.6] {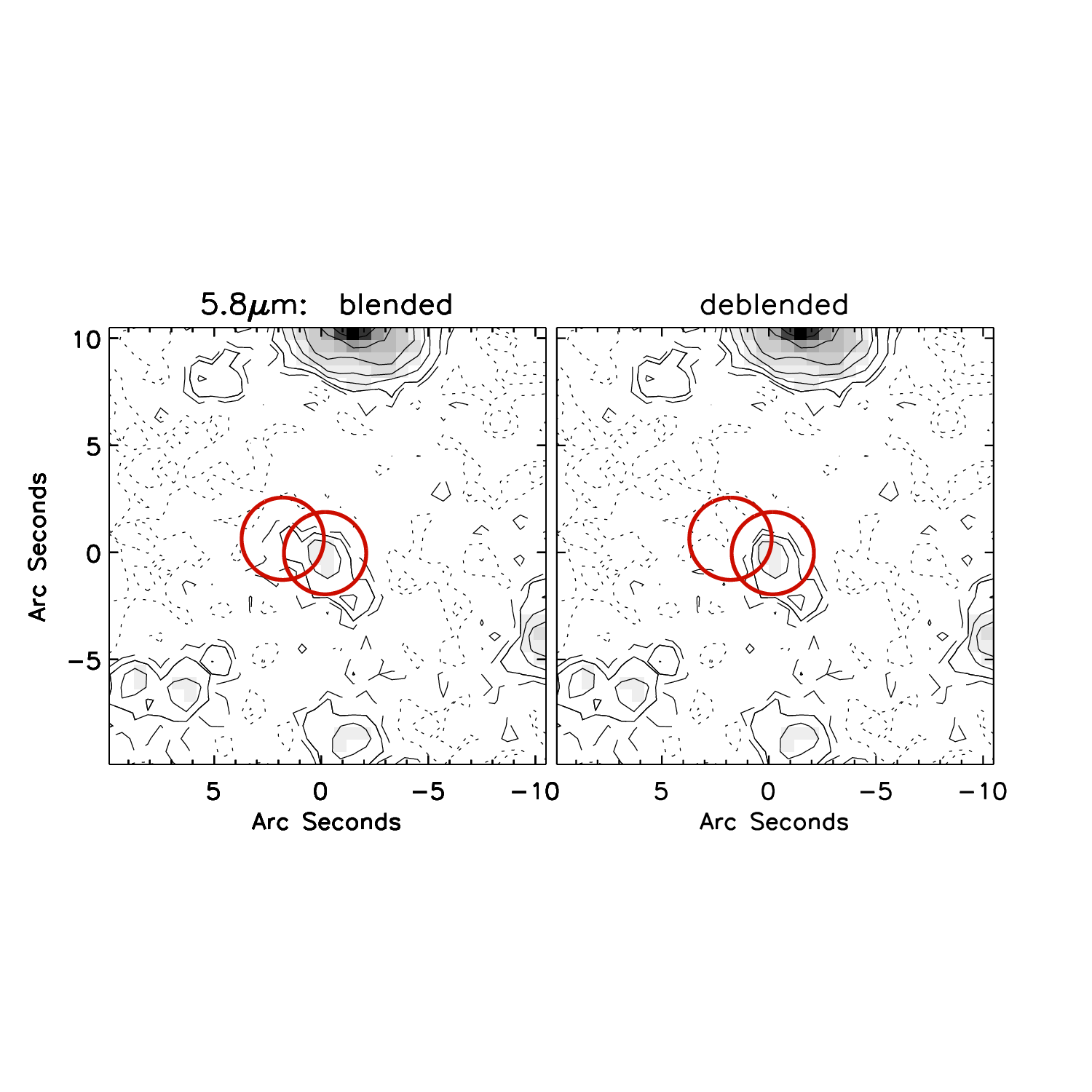}
\includegraphics[bb =0 200 432 432, scale=0.6] {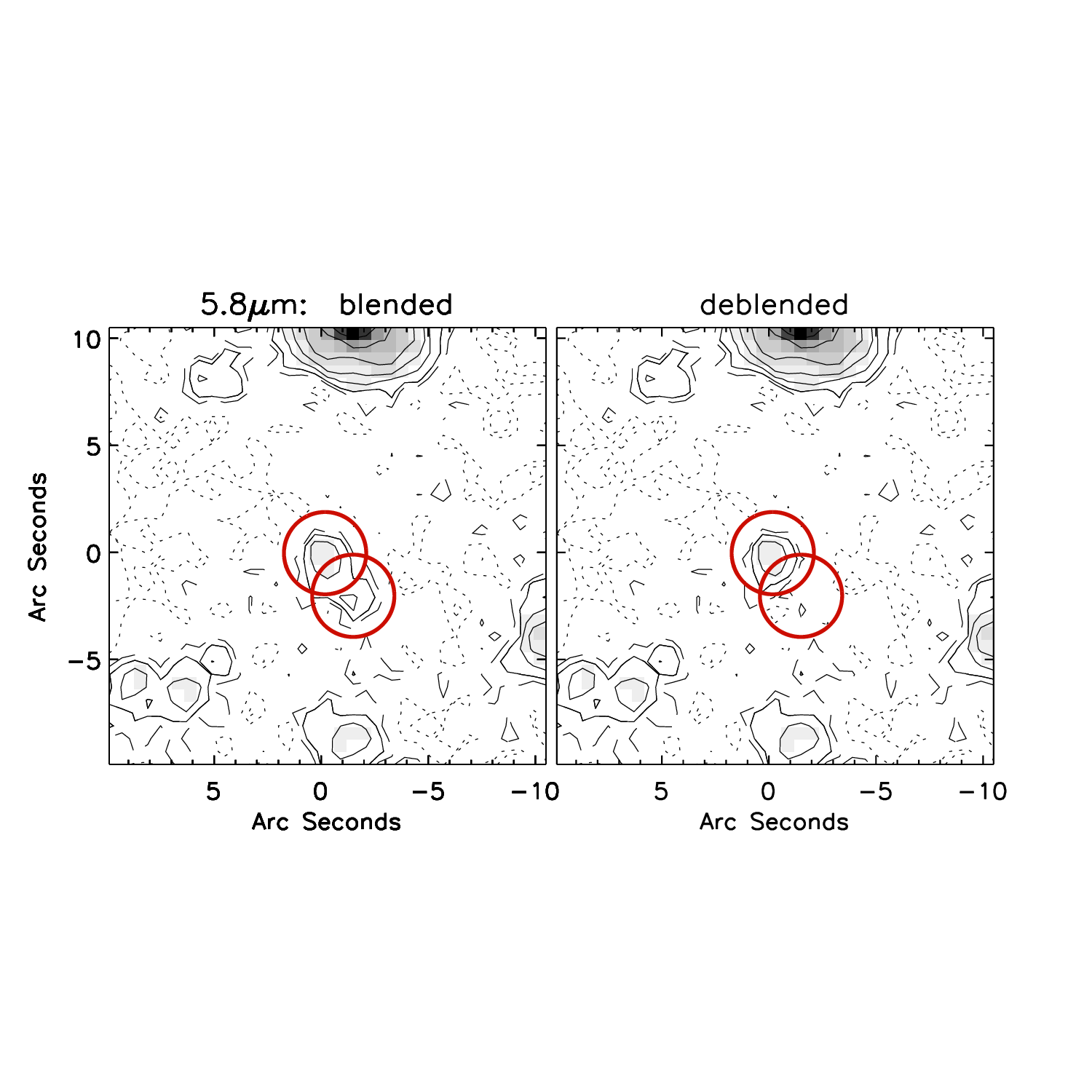}
\includegraphics[bb =0 130 432 432, scale=0.6] {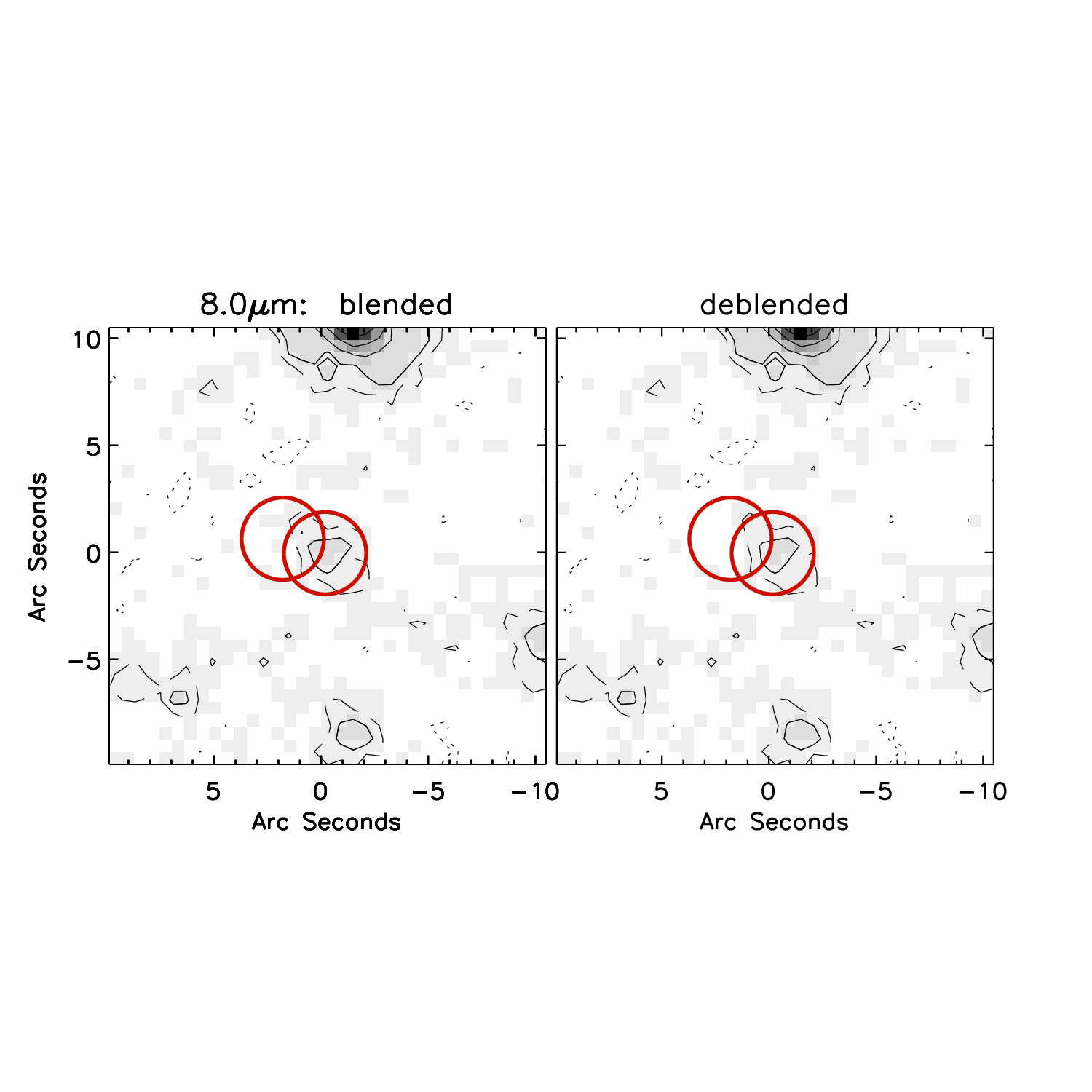}
\includegraphics[bb =0 130 432 432, scale=0.6] {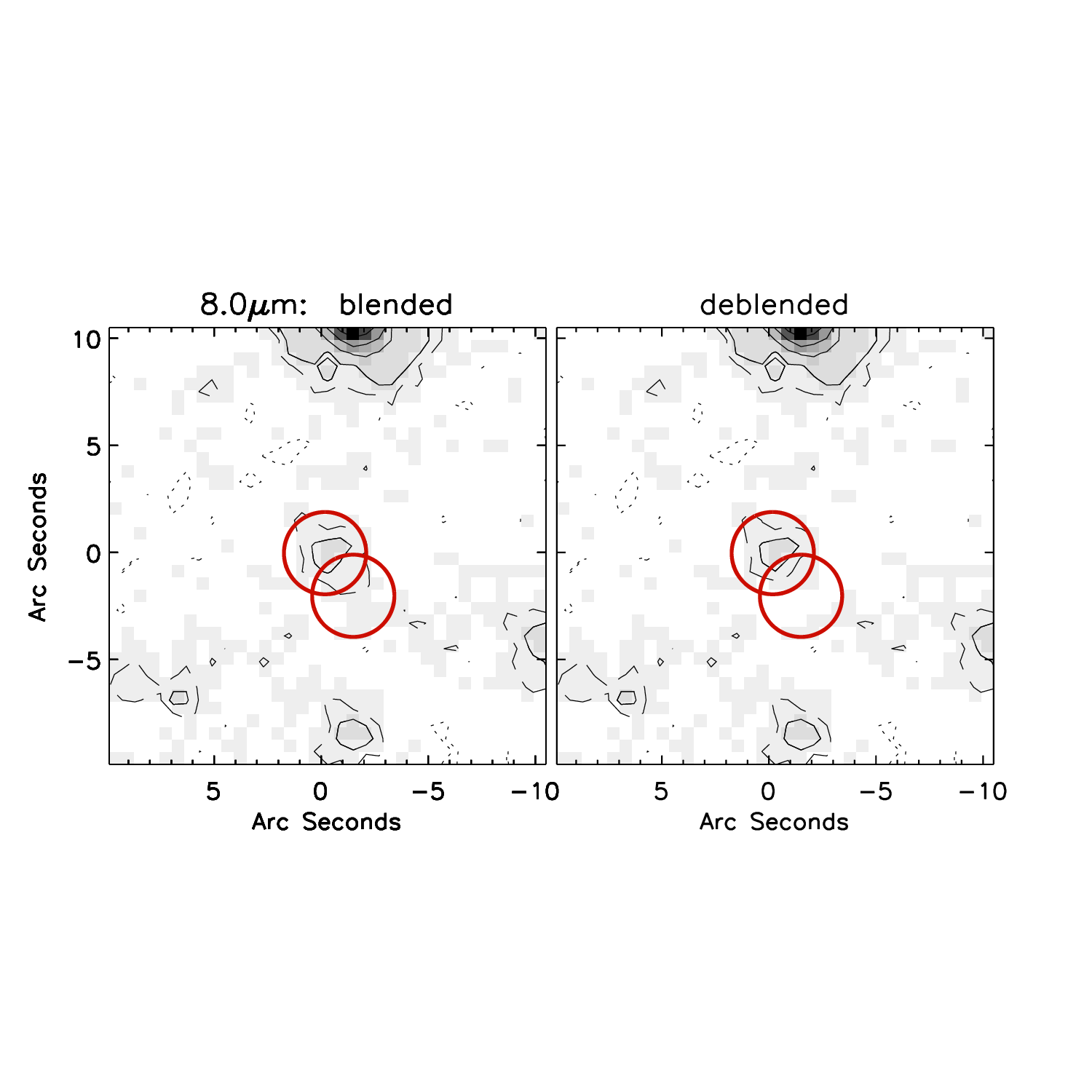}
\\ \\ \\
                \caption{Cosbo-3 (indicated by the thick circle in the center of the stamp) is
                  blended by two (NE and SW) nearby sources (also
                  indicated by circles). From top
                  to bottom the rows show IRAC 3.6, 4.5, 5.8, and 8.0
                  $\mu$m bands, respectively.  The contour levels
                  are at $-4\sigma,\, -2\sigma$ (dotted lines), $2\sigma,\,4\sigma$ (dashed lines)
                  and $2^i\sigma, i=3,4,5,6...$ (full lines), where the rms ($\sigma$)
                  has been derived locally. The deblending was done by
                  iteratively subtracting a point source (2D-Gaussian) in all four IRAC bands.}
	\label{fig:deblendc2}
\end{figure*}

\begin{figure*}[h!]
\includegraphics[bb =0 200 432 432, scale=0.6] {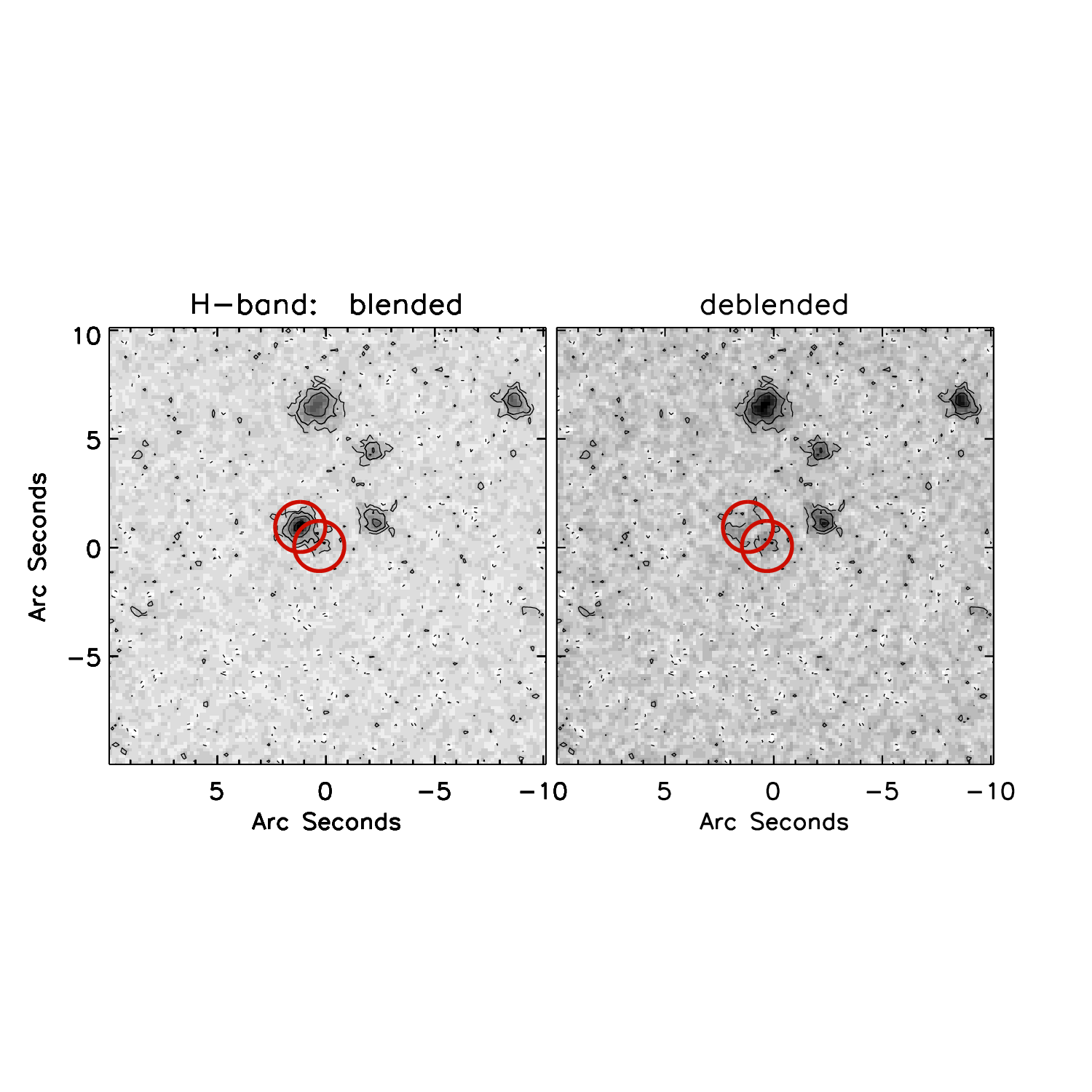}
\includegraphics[bb =0 200 432 432, scale=0.6] {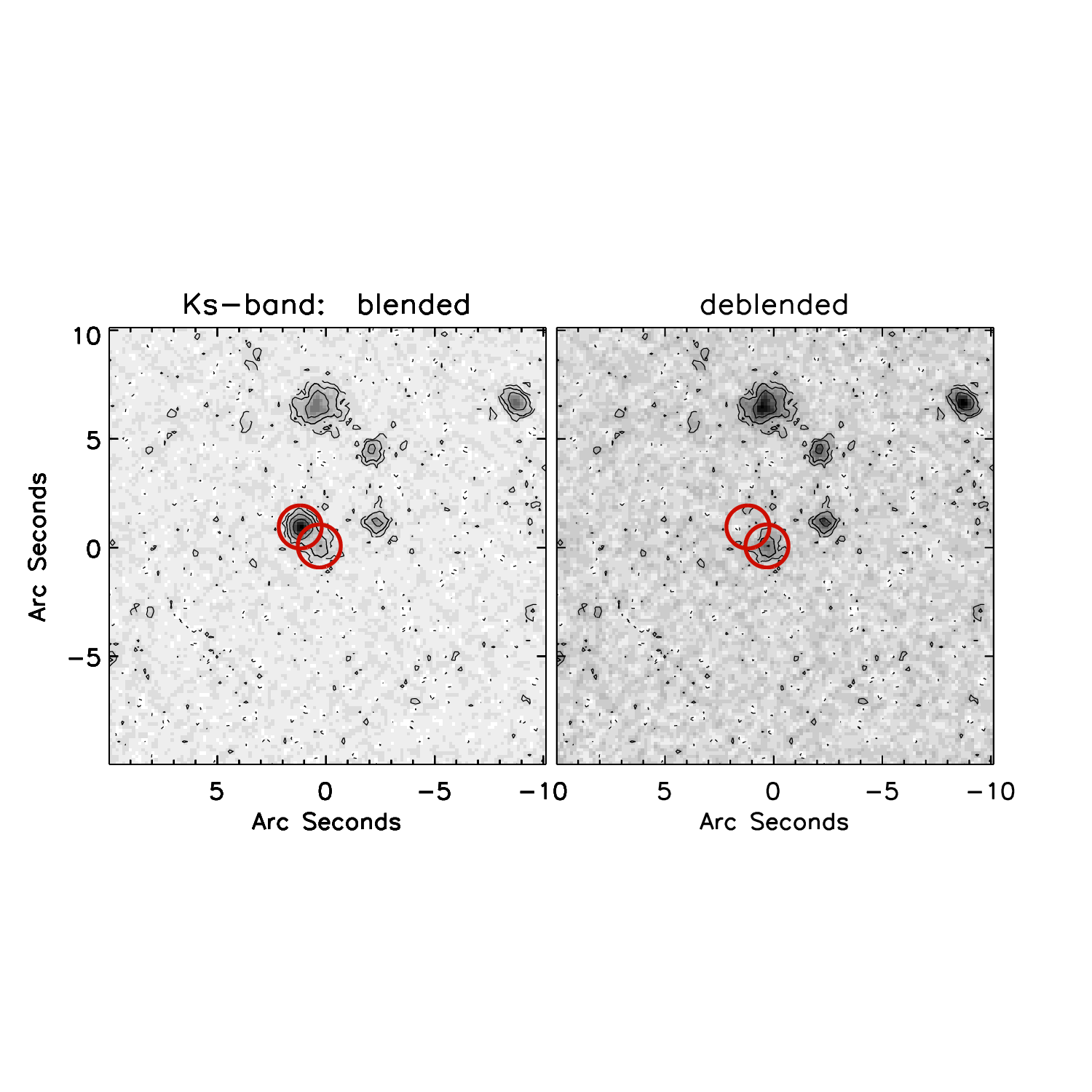}
\includegraphics[bb =0 200 432 432, scale=0.6]  {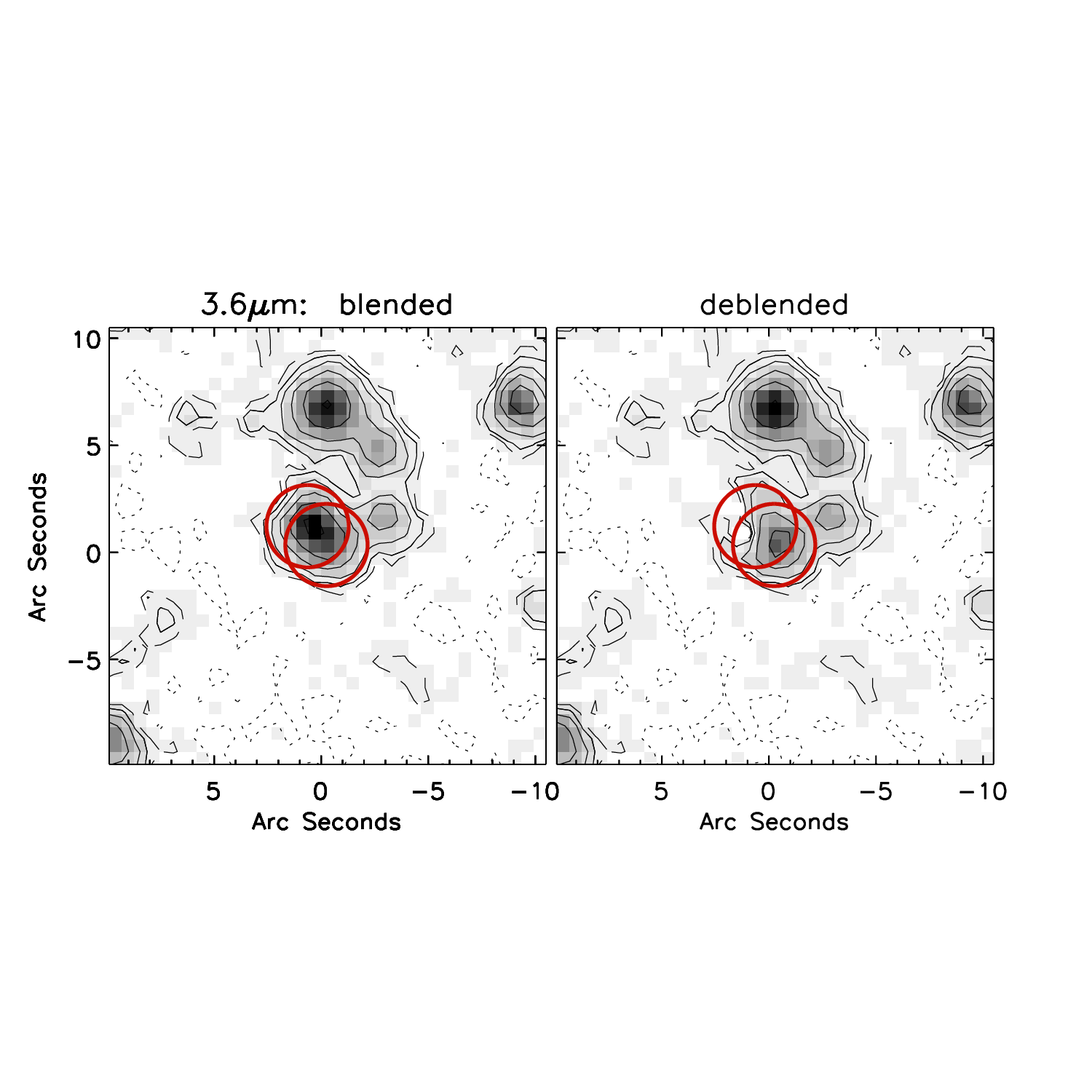}
\includegraphics[bb =0 200 432 432, scale=0.6]  {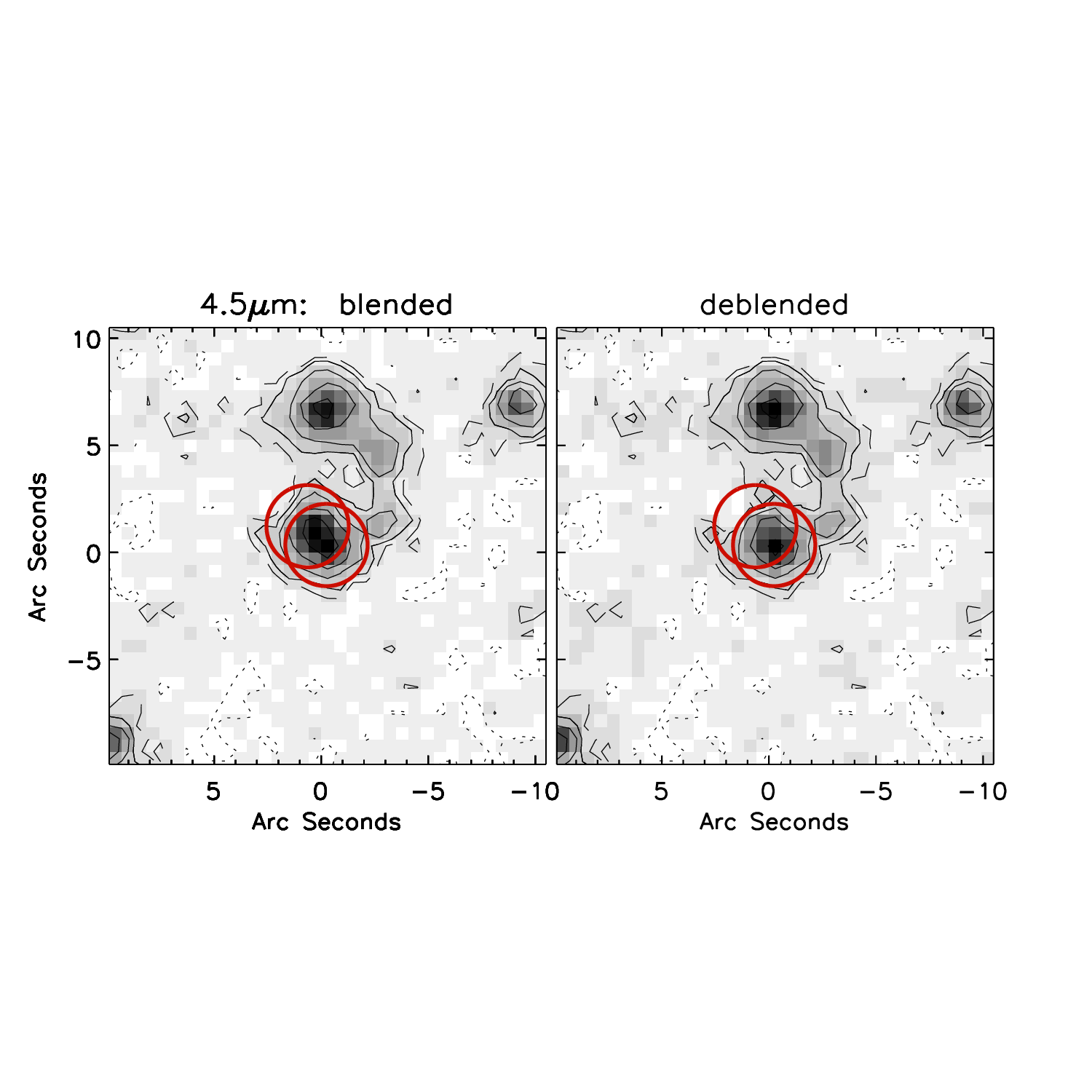}
\includegraphics[bb =0 50 432 432, scale=0.6]  {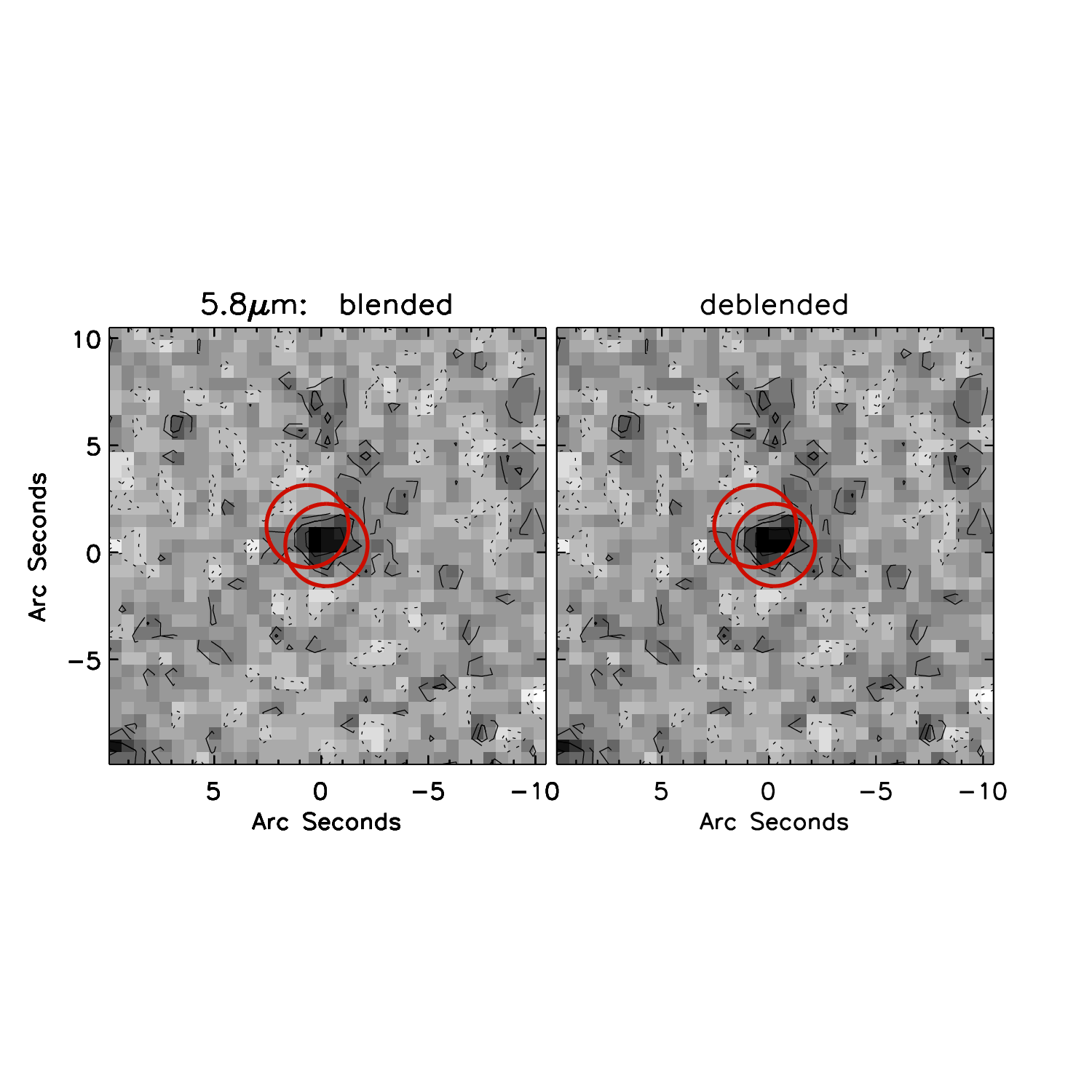}
\includegraphics[bb =0 50 432 432, scale=0.6]  {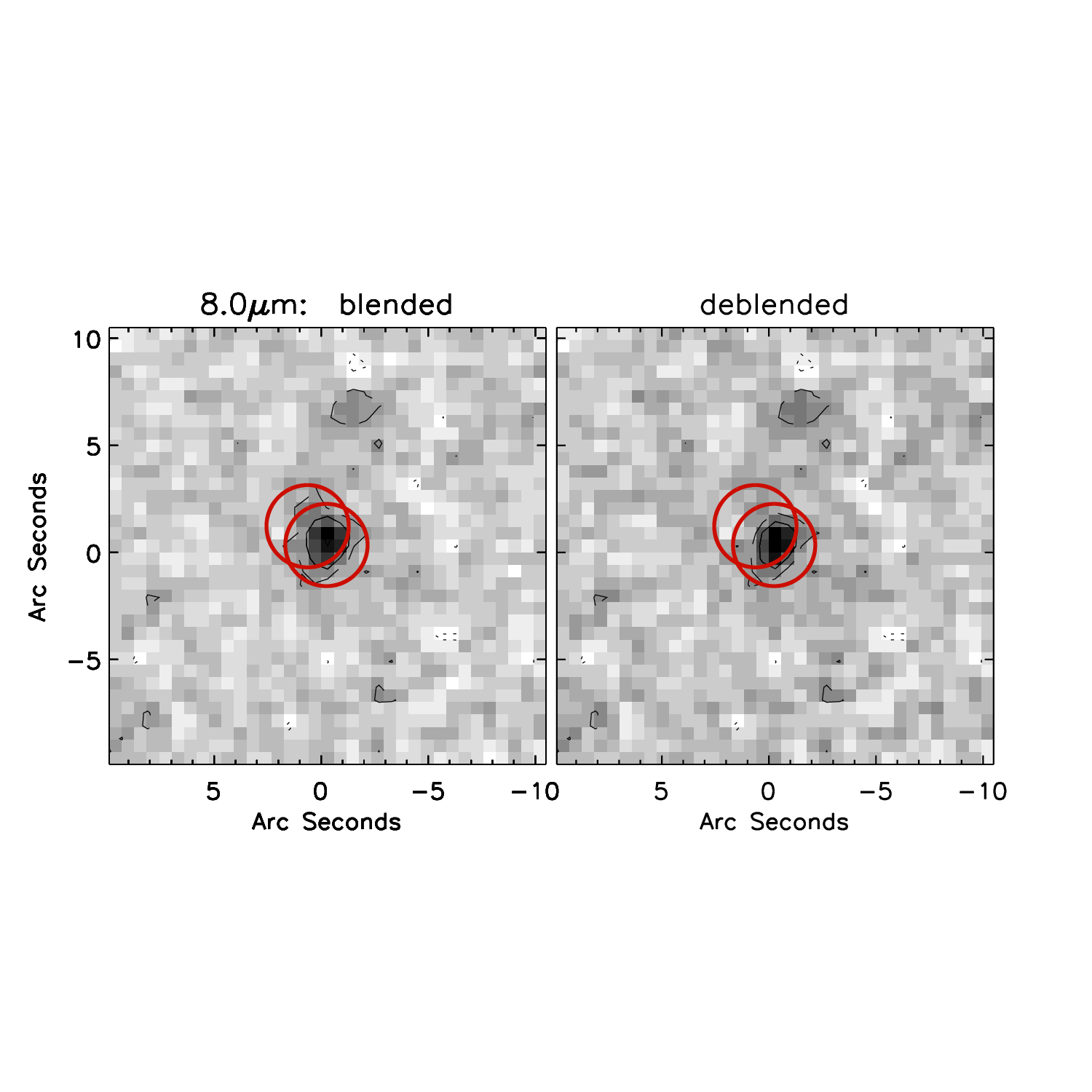}
                \caption{Blended and deblended Cosbo-8 images. H, Ks,
                  and IRAC bands (indicated in each panel) needed to
                  be deblended. The contour levels are at
                  $-4\sigma,\, -2\sigma$ (dotted lines),
                  $2\sigma,\,4\sigma$ (dashed lines) and $2^i\sigma,
                  i=3,4,5,6...$ (full lines), where the rms ($\sigma$)
                  has been derived locally. The CARMA position is
                  marked by the thick circle in the center of the stamp (its size matches
                  the photometry aperture). The source that was subtracted
                  (assuming a 2D-Gaussian point-source model) is
                  outlined by the circle to the NE. }
	\label{fig:deblendc3}
\end{figure*}

All three sources presented in this work are blended in the near/mid-IR
images. We
deblend the sources by subtracting from the corresponding image a
point source centered at the a-priori known position of the blending
source. The position was obtained either from the K$_s$ band image
(Cosbo-3 and Cosbo-8) or the pixel with maximum value in the
3.6~$\mu$m image (AzTEC/C1; see \f{fig:deblendc1} , \f{fig:deblendc2} \ and \f{fig:deblendc3} ). The peak value of the point source to be
subtracted was optimized by repeating the procedure multiple times
until a noise-level residual remained at the position of the blending
source.  The deblending uncertainty was then obtained from the
$1\sigma$ spread in the extracted magnitudes when varying the peak flux
by $\pm20\%$. This uncertainty corresponds to $\sim0.1$~mag or better
for all bands, and it is then added in quadrature to the photometry error
estimated in the previous Section in order to obtain the magnitude error.

Assuming a 2D-Gaussian point-spread function was satisfactory for
Cosbo-3 and Cosbo-8 in all affected bands (see \f{fig:deblendc2} \ and \f{fig:deblendc3} ), while the source blending AzTEC/C1 appears to
have internal structure in IRAC 3.6 and 4.5~$\mu$m images. Rather than
modeling AzTEC/C1 using a more complex (and therefore more uncertain)
model we simply isolate the blending source within an aperture of 2''
radius, mirror it around its diagonal and then subtract this from the
image (see \f{fig:deblendc1} ). This will obviously cause over-subtraction in the part of the
aperture not associated with the source of interest, while it can be
assumed that the contribution of the blending source has been well
subtracted from the source of interest. The blended and deblended
image cutouts for our sources are shown in \f{fig:deblendc1} ,
\f{fig:deblendc2} , and \f{fig:deblendc3} ,  and the extracted photometry for each source is given in \t{tab:phot} .

\subsection{Photometric redshifts}
\label{sec:photz}

In this Section we first investigate photometric redshift estimates for
SMGs based on a sample of eight interferometrically observed SMGs with
spectroscopic redshift spanning a broad redshift range of $z\sim1-5$ present in the COSMOS field (\s{sec:calibphotz} ). Showing that photometric redshifts can be reliably calculated for SMGs, we then derive photometric redshifts using the same method for the three CARMA-COSMOS SMGs analyzed here (\s{sec:redsft} ).

\subsubsection{Calibration of photometric redshifts for SMGs}
\label{sec:calibphotz}

Photometric redshifts are computed by fitting optimized spectral templates to
the spectral energy distribution of a given galaxy, leaving redshift
as a free parameter. The redshift is then determined via a $\chi^2$
minimization procedure. We use the Hyper-z code to compute photometric
redshifts for our SMGs with the same parameterization as in
Wardlow et al.\ (2010, 2011): Extinction assuming a Calzetti et al.\
(2000) law, with reddening ($A_V$) varied from 0 to 5, and an allowed
redshift range of 0 to 7. Based on $\sim30$ SMGs with spectroscopic
redshifts, drawn from the LESS survey, Wardlow et al.\ (2011) have
shown that photometric redshifts for SMGs derived with Hyper-z using
the above mentioned parameterization are estimated accurately (they
find a median offset between the spectroscopic and
photometric redshifts of $0.023\pm0.021$; see also their Fig.~1).

In general the quality of photometric-redshifts will depend on the
choice of the spectral template library to be fit. Therefore, in
addition to the templates provided by Hyper-z (similar to those used by Wardlow
et al.; 6T hereafter; see below), we also test other sets of template
libraries. Our spectral
model libraries are summarized as follows:
\begin{itemize}
\item[{\bf 2T}:] Only two -- burst and constant star formation
  history -- templates drawn from the Bruzual \& Charlot (2003)
  library (and provided with Hyper-z).
\item[{\bf 6T}:] Six templates provided by the Hyper-z code: burst, four
  exponentially declining star formation histories (star formation
  rate $\propto e^{-t/\tau}$ where $t$ is time, and $\tau=0.3,\, 1,\,
  3$~and~5~Gyr) and a constant star formation history. This selection of SFH/templates is similar to the approach used by Ilbert et al. (2010) to compute stellar masses with LePhare.
\item[{\bf M}:] Spectral templates developed in GRASIL (Silva et al.\
  1998; Iglesias-P\'{a}ramo et al.\ 2007) and optimized for SMGs by Michalowski et al.\ (2010).
\end{itemize}

For this analysis we use eight SMGs in the COSMOS field with available
spectroscopic redshifts of counterparts determined interferometrically
(Capak et al.\ 2008, 2010; Schinnerer et al.\ 2008; Riechers et al.\
2010; \smo\ et al.\ 2011; Karim et al., in prep, Sheth et al., in
prep). We compute the photometric redshifts
for these SMGs as described above and show the total $\chi^2$
distribution as a function of redshift in \f{fig:photz} . The overall
match between the most probable photometric redshift (corresponding to
the minimum $\chi^2$) and the spectroscopic redshift is highly
satisfactory. There are no catastrophic outliers. For source AzTEC-3
at $z_\mathrm{spec}=5.3$ (Capak et al.\ 2010; Riechers et al.\ 2010)
there are two $\chi^2$ minima. However, the low redshift peak can be
disregarded given that the galaxy is not detected in the radio. If it
were a low-redshift SMG one would expect a strong radio detection
given the depth of the VLA-COSMOS survey.

Overall, all templates reach similar solutions, and the best agreement (i.e.\ tightest $\chi^2$ distribution)
between the spectroscopic and photometric redshifts is reached when
using the Michalowski et al.\ (2010) spectral templates, and hereafter we adopt
these  for our photometric-redshift computation.  A direct
comparison between the photometric (based on M templates) and
spectroscopic redshifts is given in \f{fig:spec-phot-z} . The errors
indicate the 99\% confidence interval.  We find a median of -0.03, and
a standard deviation of 0.08 in the
$(z_\mathrm{phot}-z_\mathrm{spec})/(1+z_\mathrm{spec})$
distribution. We conclude that our photometric redshift computation is
accurate for SMGs as expected based on results from previous studies
(e.g.\ Daddi et al.\ 2009; Wardlow et al.\ 2010, 2011; Yun et al.\
2011).

\subsubsection{Redshifts for CARMA-COSMOS SMGs}
\label{sec:redsft}

\begin{figure*}
\includegraphics[bb = 90 430 486 652,scale=0.45]{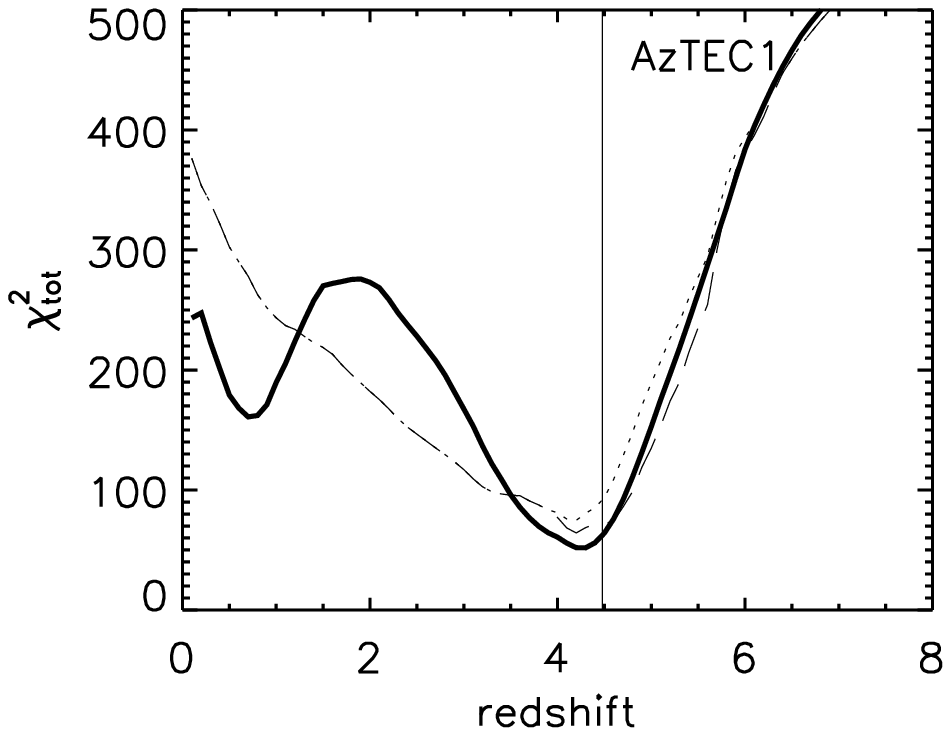}
\includegraphics[bb = 210 430 486 602,scale=0.45]{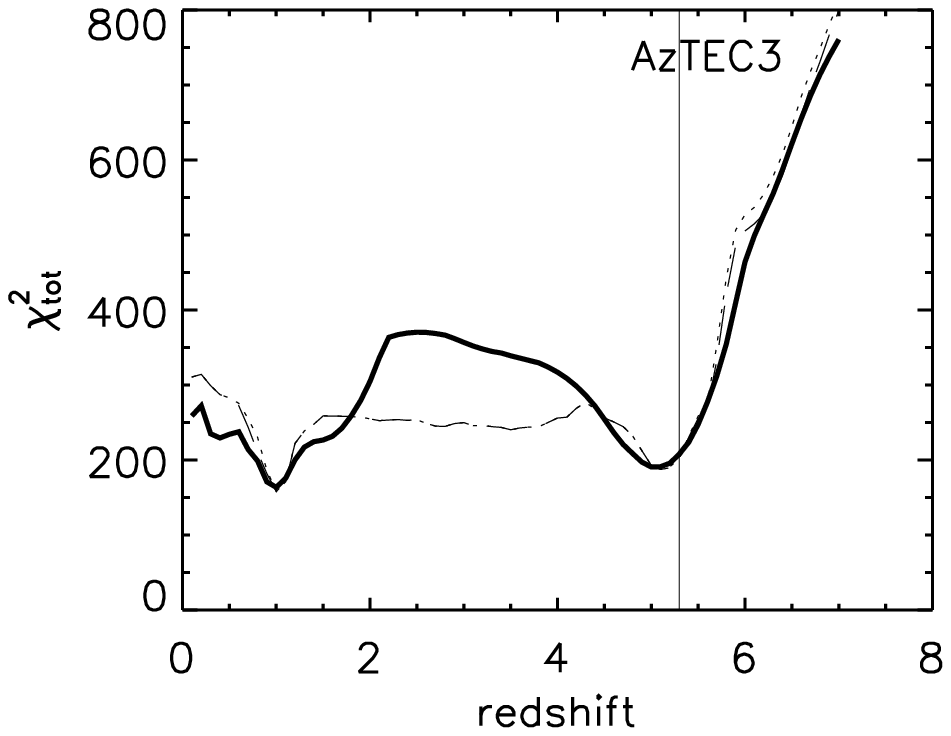}
\includegraphics[bb = 210 430 486 602,scale=0.45]{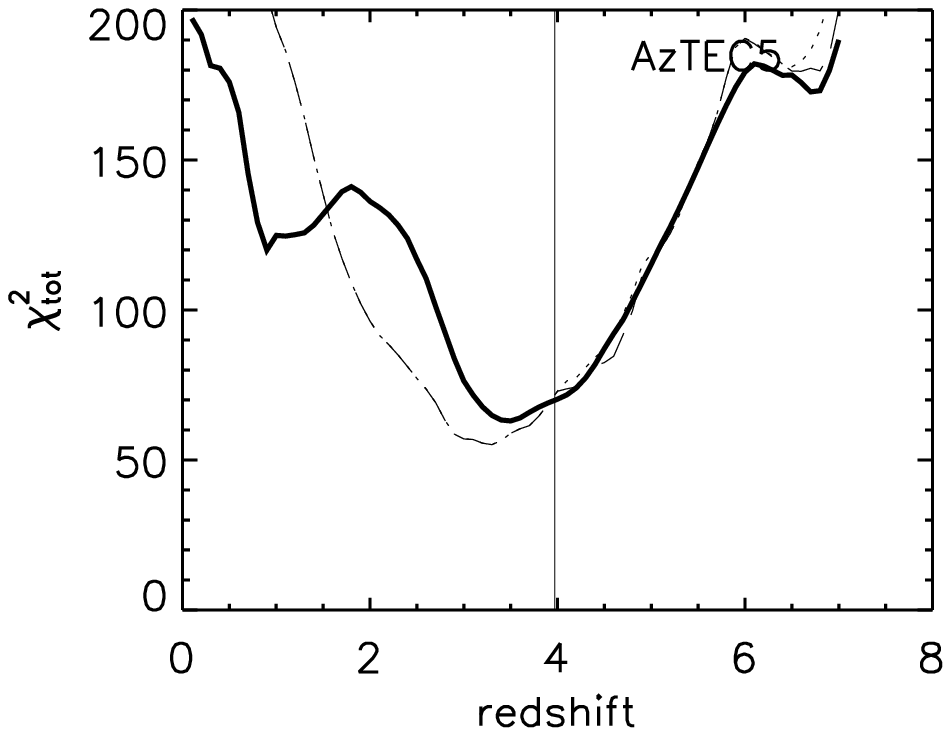}
\includegraphics[bb = 210 430 256 602,scale=0.45]{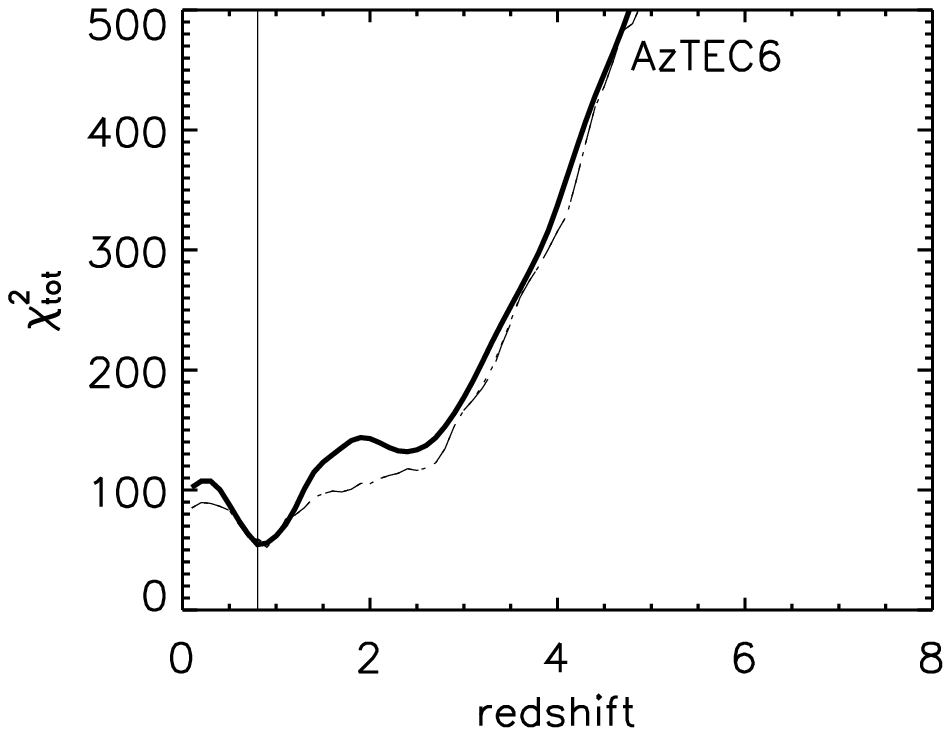}\\
\includegraphics[bb = 90 410 486 652,scale=0.45]{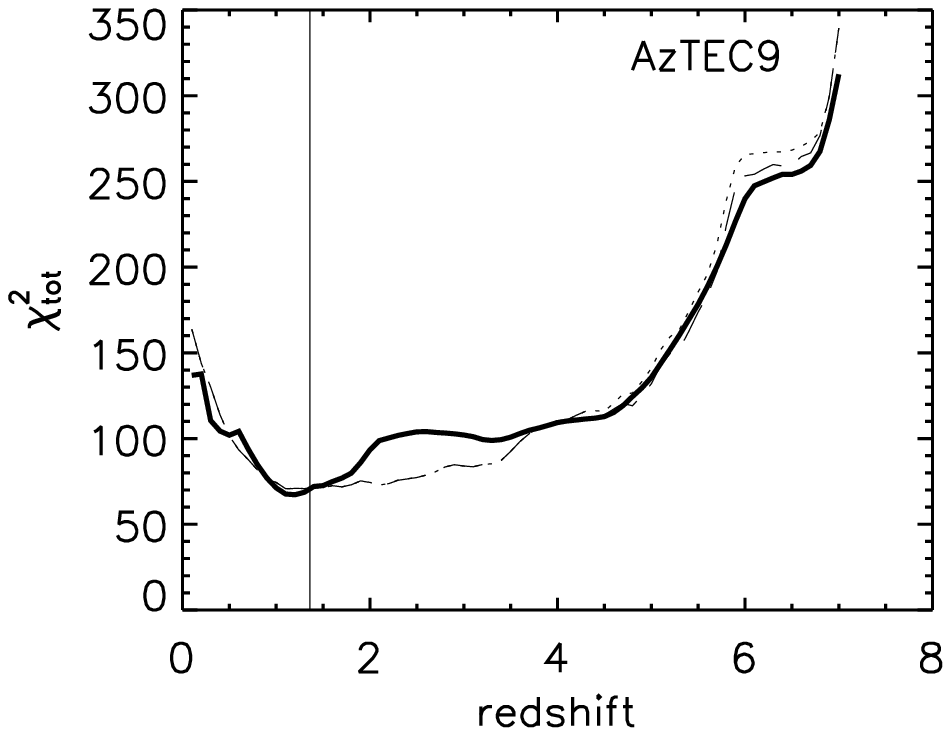}
\includegraphics[bb = 210 410 486 652,scale=0.45]{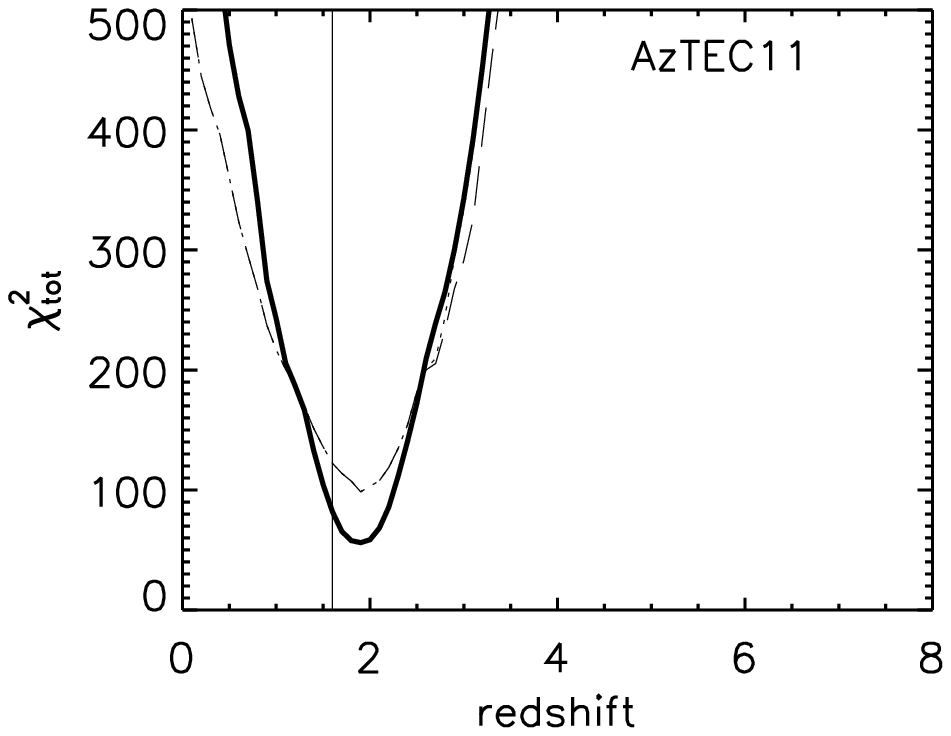}
\includegraphics[bb = 210 410 486 652,scale=0.45]{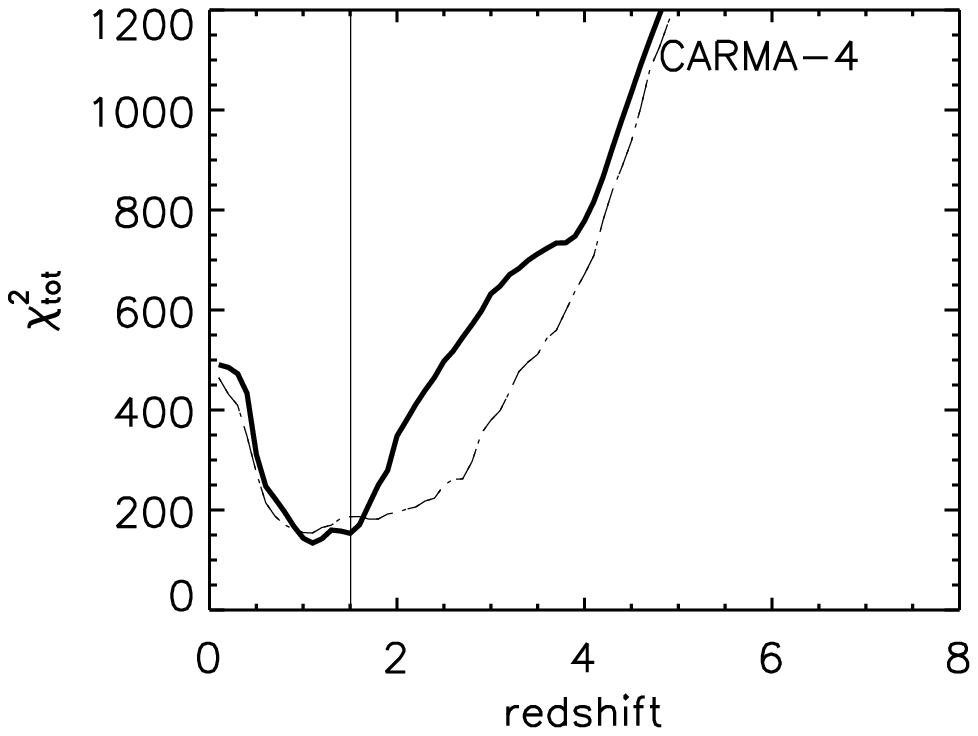}
\includegraphics[bb = 210 410 256 652,scale=0.45]{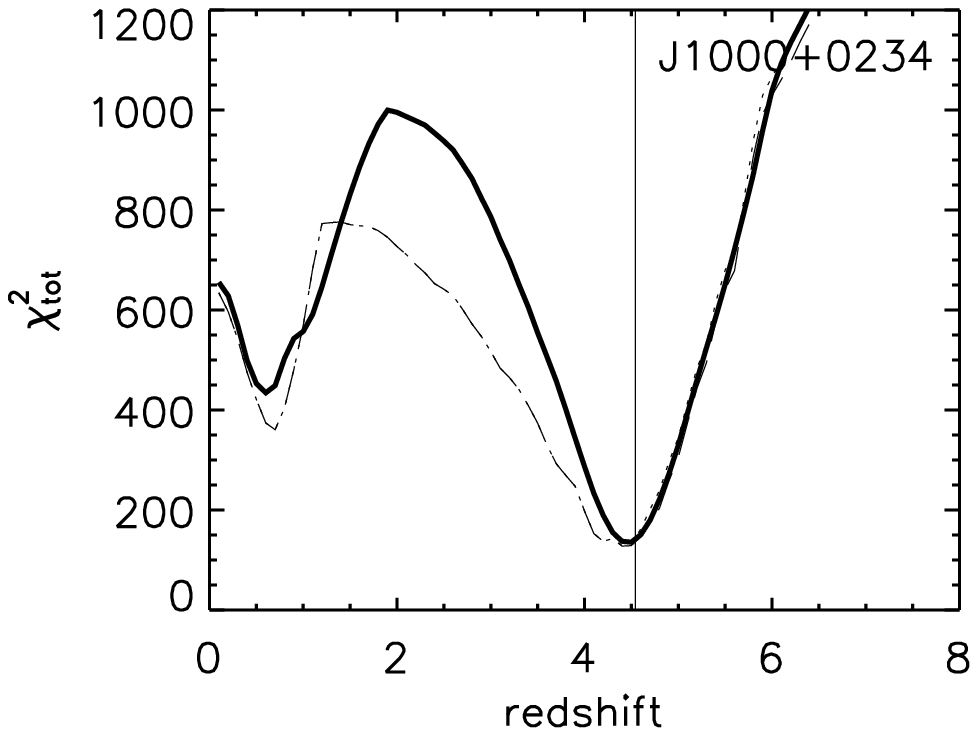}
\caption{Full $\chi^2$ distribution of the photometric redshift
  determination for SMGs in the COSMOS field with interferometrically
  determined counterparts with spectroscopic redshifts. The
  photometric redshifts were derived using various sets of spectral
  models (see text for details): 2T (dotted lines), 6T (dashed-lines),
  M (full lines). The spectroscopic redshifts are indicated by
  vertical lines. Note that for AzTEC-3 the radio 
  non-detection rules out the first $\chi^2$ minimum.}
      \label{fig:photz}
\end{figure*}

\begin{figure}
\includegraphics[bb =  54 410 486 792,width=\columnwidth]{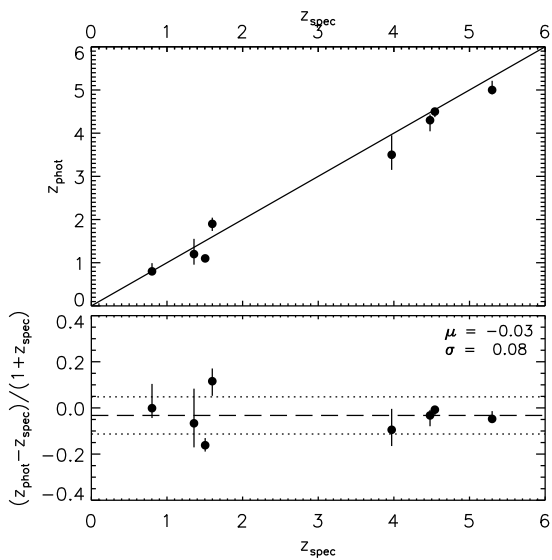}
\caption{Comparison of spectroscopic and photometric redshifts for
  eight SMGs from the COSMOS field using the Michalowski et al.\ (2010)
  spectral templates. The errors show the $99\%$ confidence
  interval. The median offset and standard deviation of the $\Delta z
  / (1+z_\mathrm{spec})$ distribution are indicated in the bottom
  panel.  }
      \label{fig:spec-phot-z}
\end{figure}

\begin{figure*}
\includegraphics[bb=80 500 486 682, scale=1.1]{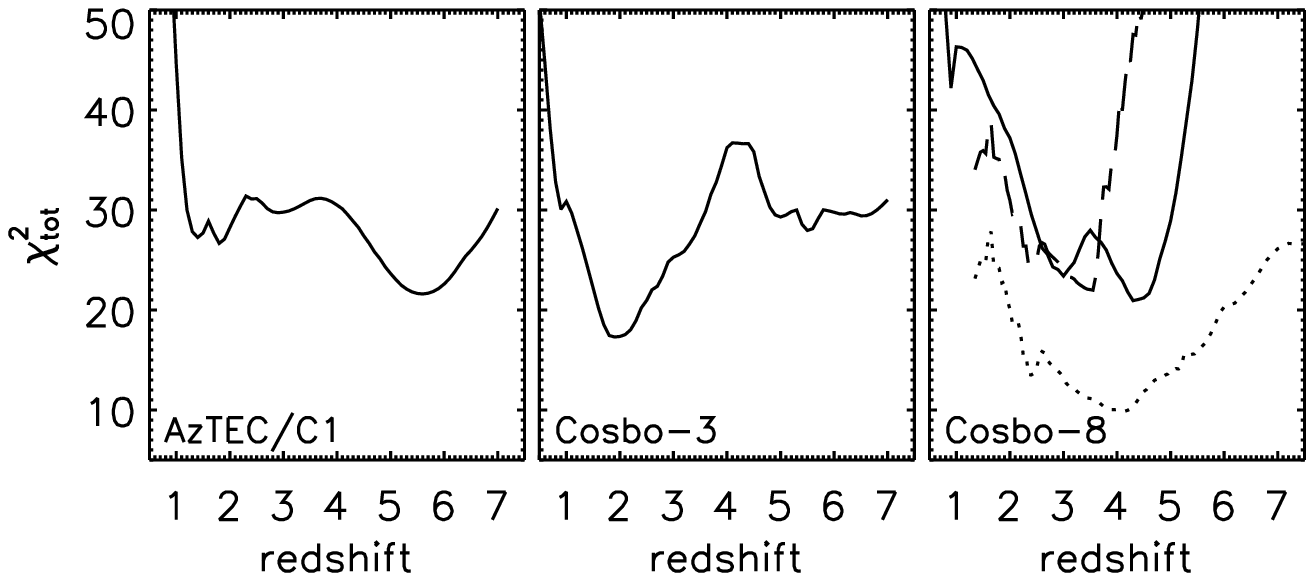}\\
\includegraphics[bb=80 490 486 680, scale=1.1]{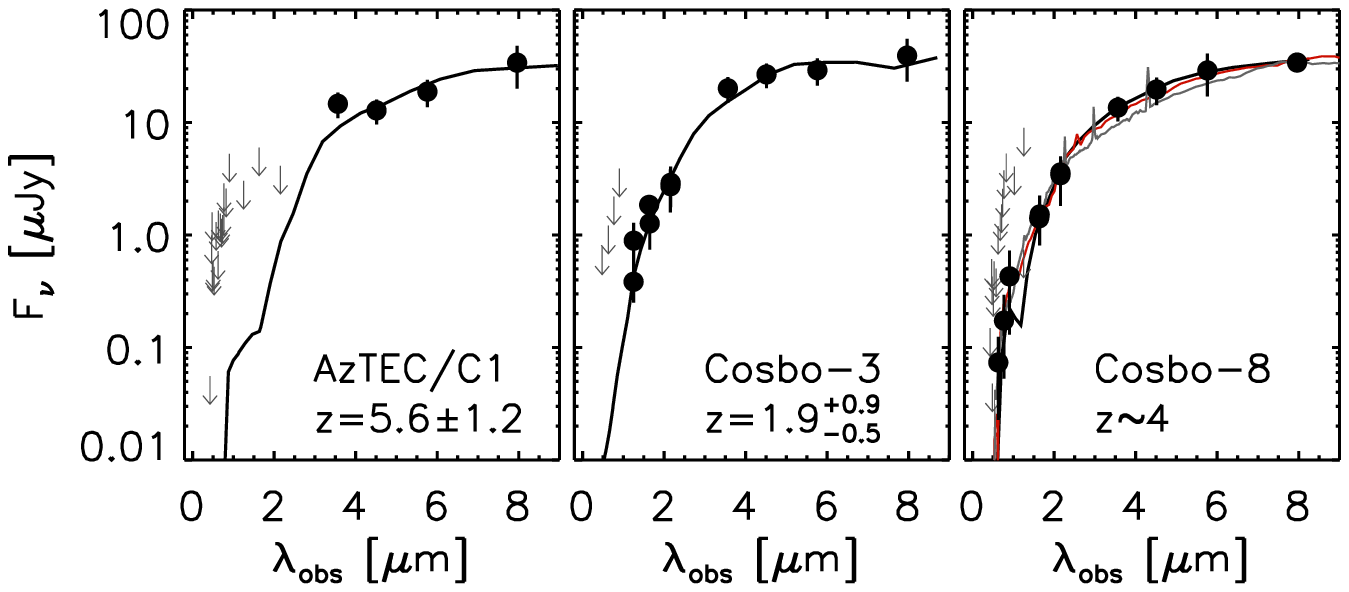}
\caption{ Top panel: Total $\chi^2$ distribution as a function of redshift for our
  SMGs (indicated in the panels).  Solid curves show the solution based on the Michalowski et al.\ (2010) spectral model library used to derive the photometric redshifts  as described in \s{sec:photz} . For Cosbo-8 (right panel) we also show the total $\chi^2$ distribution when using model libraries with AGN templates and a low-luminosity AGN prior (dashed line) and a QSO prior (dotted line) are used (see text for details).  Bottom panel: The spectral energy distribution for our SMGs (indicated in the panels; dots and arrows for $5\sigma$ upper limits) with the best fit spectral model (corresponding to that yielding the minimal $\chi^2$ value) overplotted (black, gray and red lines show the best fit Michalowski et al., AGN and QSO library spectra, respectively). The best fit redshift is also indicated. 
  }
	\label{fig:photoz}
\end{figure*}

\begin{table*}
\begin{center}
\caption{Multi-wavelength photometry}
\label{tab:phot}
\vskip 10pt
\begin{tabular}{|c|c|c|c|c|}
\hline
    Telescope/Band &  \multicolumn{3}{|c|}{AB magnitude} \\
          &  AzTEC/C1  &  Cosbo-3 & Cosbo-8  \\
\hline  
Subaru/g$^+$ & $>26.5$  & $>26.5$ & $>26.5$\\
Subaru/r$^+$ & $>26.5$ & $>26.5$& $26.8\pm0.3$\\
Subaru/i$^+$  & $>26.1$ & $>26.1$& $25.8\pm0.3$\\
Subaru/z$^+$  &$>25.1$ & $>25.1$ &$24.8\pm0.3$\\
UltraVista/J     & $>23.9$ & $24.15\pm0.19$ & $>23.9$ \\ 
UltraVista/H        &$>22.7$ & $23.64\pm0.18$ & $23.44\pm0.20$ \\
UltraVista/K$_s$    &$>22.4$ & $22.80\pm0.18$& $22.50\pm0.20$\\ 
Spitzer/3.6~$\mu$m & $21.12\pm0.11$ & $20.78\pm0.11$ & $21.20\pm0.11$ \\
Spitzer/4.5~$\mu$m & $21.28\pm0.11$ & $20.48\pm0.11$ & $20.81\pm0.11$ \\
Spitzer/5.8~$\mu$m & $20.83\pm0.12$ & $20.35\pm0.12$ & $20.36\pm0.12$ \\
Spitzer/8.0~$\mu$m & $20.12\pm0.18$ & $19.96\pm0.18$ & $20.12\pm0.18$\\
JCMT/AzTEC-1.1~mm & $13.62^{+0.09}_{-0.08}$ & $13.94^{+0.12}_{-0.11}$ & $14.98^{+0.32}_{-0.35}$ \\
IRAM 30-m/MAMBO-1.2~mm & -- &$14.22\pm0.16$ & $14.56\pm0.20$\\
CARMA-1.3~mm & $14.23\pm0.34$ & $14.57\pm0.14$& $14.70\pm0.34$ \\
VLA-20~cm & $19.80\pm0.25$ & $19.17\pm0.19$ & $18.86\pm0.14$ \\
\hline 
\end{tabular}
\end{center}
Magnitudes are total magnitudes already corrected for reddening.
Limits are 
either adopted from \citet[][
g$^+$, r$^+$, i$^+$, z$^+$]{capak07} or extracted from the aperture flux
(J, H, K$_s$). 
\vspace{1cm}
\end{table*}

\begin{table*}
\begin{center}
\caption{Best fit properties given by the photometric redshift computation}
\label{tab:sed}
\vskip 10pt
\begin{tabular}{ccc}
\hline
    Source        & template &  Hyper-z $A_V$$^+$ \\\hline 
      AzTEC/C1   &     SMMJ030226.17+000624.5 &  0.00    \\
      Cosbo-3	&     LESSJ033229.4-275619       & 2.55 \\
      Cosbo-8$^a$	&     SMMJ131215.27+423900.9  & 3.00 \\
\hline
\end{tabular}
\end{center}
$^{+}$ Reddening computed by Hyper-z; note that the templates already have intrinsic reddening as defined by Michalowski et al.\ (2010; see their Tab.~A3) \\
$^a$ Best fit template/$A_V$ for both $\chi^2$ minima. 
\end{table*}

  Following the same approach as described in the previous Section we compute the photometric redshifts for
  AzTEC/C1, Cosbo-3 and Cosbo-8 using their extracted (deblended)
  photometry (see \t{tab:phot} ). The results are presented in
  \f{fig:photoz} \ and \t{tab:sed} . We find photometric redshifts of $5.6\pm1.2$, and
  $1.9^{+0.9}_{-0.5}$ for AzTEC/C1, and Cosbo-3, respectively. We
  stress that our photometric redshift for Cosbo-3 is consistent (within $\Delta z=0.5$) with the recently confirmed spectroscopic redshift (Riechers
  et al., in prep), affirming both our computation of the UV-MIR-based
  photometric redshifts, and deblending technique.  
  
  Cosbo-8 is detected in the X-rays within the Chandra-COSMOS survey ($F_\mathrm{0.5-10~keV}=(2.49\pm 0.65)\times10^{-15}~\mathrm{erg/cm^2/s}$) suggesting the presence of an AGN. As AGN are variable sources with featureless power-law spectral energy distributions, special treatment for photometric redshift estimates is required. In the COSMOS survey Salvato et al.\ (2009, 2011) have optimized the photometric redshift computation for X-ray selected sources reaching an accuracy of $\sigma_\mathrm{\Delta z/(1+z_{spec})}\sim0.015$. 
 Salvato et al.\ (2011) find that the Chandra-COSMOS source associated with Cosbo-8 (CID 838; see Civano et al 2012 submitted) has a photometric redshift corresponding to $z_\mathrm{phot}=0.82 \pm0.02$ best fit by a normal galaxy template. However, the photometry is contaminated by the presence of a nearby object.
Deblending and extracting the photometry as described here,  with the addition of  J, H, Ks from UltraVista, the photometric redshift value ranges from $z_\mathrm{phot} =3.6 - 4.3$, depending on the luminosity prior adopted (i.e. assuming a luminosity typical of a low luminosity AGN, i.e.\ $-8<M_B<-24$, or typical of a QSO, i.e.\ $-20<M_B<-30$, respectively; Veron \& Veron 1998). Assuming a low luminosity AGN the best fit template is a Seyfert~1.8 drawn from the Polletta et al. (2007) library while assuming a luminosity typical of QSOs, the best fist template is a hybrid created using a ULIRG (IRAS22491) and a QSO (see \f{fig:photoz} \  and Salvato et al 2009 for details).
The solutions are consistent within $1\sigma$ as in both cases the redshift probability distribution function shows a broad range of possible solutions, rather than a well defined unique peak.  These solutions are also consistent with the photometric redshift value computed as described in the previous section which yields two $\chi^2$ minima (at $z\sim3$  and $z\sim4$; see \f{fig:photoz} ). Comparing the results from the various spectral libraries the best $\chi^2$ value is obtained when using a QSO prior yielding a redshift of $z_\mathrm{phot}=4.1^{+0.2}_{-0.5}$ (where the error is a $1\sigma$ error; see  \f{fig:photoz} ). Hereafter we take $z\sim4$ as the best redshift estimate for this source, noting that spectroscopic redshift follow-up is required to disentangle between the various photometric redshift solutions for this SMG.
    
For comparison, the mm-to-radio flux ratios of the sources, regularly utilized as a
redshift estimate for SMGs (Carilli \& Yun 2002), suggest that all
three sources lie at high ($z\gtrsim2$) redshift (see \t{tab:det}
). An improved version of the dust-independent Carilli \& Yun (2002)
redshift estimator via the observed mm-to-radio flux density ratio
(Yun et al.\ 2011) yields redshifts of $4.3^{+0.7}_{-1.4}$,
$3.2^{+0.6}_{-1.0}$, and $1.9^{+0.5}_{-0.7}$ for AzTEC/C1, Cosbo-3,
and Cosbo-8, respectively (using AzTEC 1.1~mm measurements for all sources).

  In summary, we find photometric redshifts of $5.6\pm1.2$,
  $1.9^{+0.9}_{-0.5}$, and $\sim4$ for AzTEC/C1, Cosbo-3, and
  Cosbo-8, respectively.   Below we summarize the
  properties of each SMG.

\subsection{Properties of individual sources}

\subsubsection{AzTEC/C1}

AzTEC/C1 has neither a J, H or Ks band counterpart
, while it can be associated with a source at
3.6~$\mu$m (and becoming most prominent at 8.0~$\mu$m) that is, however,
strongly blended with a bright source $1.97"$ to the NE (see
\f{fig:stamps} \ and \f{fig:deblendc1} ). The
radio-counterpart of AzTEC/C1 is only 0.28'' away from the reported
CARMA detection.\footnote{Based on the $\mathrm{S/N}=4.4$ radio detection at a
resolution of $\Theta=1.5"$ the expected astrometric accuracy is
$\Theta/(\mathrm{S/N})=0.34"$ (note that the overall VLA-COSMOS
astrometric accuracy is estimated to be better than $0.130"$; see
\citealt{schinnerer07} for details). In order to assess the
astrometric accuracy of our CARMA detection we imaged 3C273, our
secondary calibrator $\sim37\deg$ away from the AzTEC/C1 field phase
center. We recover its position within $0.11''$ of the nominal
position. This yields that the positional uncertainty of AzTEC/C1 is
likely better than this value, and thus it rules out the possibility
that the bright IR galaxy $\sim2"$ away from AzTEC/C1 (which also
corresponds to the closest source to AzTEC/C1 detected in the optical)
is its counterpart (or that of the radio source).} The deblended
images and photometry for AzTEC/C1 are presented in \f{fig:deblendc1}
\ and
\t{tab:phot} .

The mm-to-radio flux ratio of AzTEC/C1 yields a redshift of
$z_\mathrm{mm/radio}=4.3^{+0.7}_{-1.4}$, consistent with the
photometric redshift of $z_\mathrm{phot}=5.6\pm1.2$ derived from its
UV-MIR photometry. The inferred high redshift is consistent with the
source not being detected at wavelengths shorter than 3~$\mu$m. At
such a redshift (4.3-5.6) the radio flux density of the galaxy
($\mathrm{F_{20cm}}=44\pm10~\mu$Jy) would imply a 20~cm luminosity of
$(6-10)\times10^{24}$~\wh . If the entire radio emission arises from
star formation in the galaxy, and if at these redshifts locally
determined radio-star formation rate calibrators (Bell et al.\ 2003,
Yun et al.\ 2001) can be applied (as would be suggested by the
constancy of the FIR-radio correlation out to high redshifts; Sargent
et al.\ 2010a, 2010b, Murphy 2009), this radio luminosity would imply
a SFR of $\sim3200-5600$~\msolyr . This is somewhat in excess of
expectations for typical SMGs, thus it may be possible that part of
the (radio) emission from this source arises from black hole
accretion. However, it is worth noting that such properties are not
unusual for $z>4$ SMGs. For example, the properties of AzTEC/C1 are
very similar to those of AzTEC-1 -- the brightest SMG in the AzTEC/JCMT
COSMOS survey (Scott et al.\ 2008; Younger et al.\ 2007, 2009, \smo\
et al.\ 2011; $\mathrm{F_{20cm}} = 42~\mu$Jy, $z=4.6$).

\subsubsection {Cosbo-3}

A source coincident with the position of Cosbo-3 is
detected in J-band, as well as in longer wavelength bands. In the
Spitzer images two surrounding sources ($1.9"$ to the NE and $2.4"$ to
the SW, respectively) are blending its IR emission (see
\f{fig:stamps} \ and \f{fig:deblendc2} ).  The deblended
images and photometry for AzTEC/C1 are presented in \f{fig:deblendc2}
\ and
\t{tab:phot} .

Within the MAMBO 11'' beam there are 2 radio sources present, at separations of 
$1.3''$ and $5.9''$, respectively. 
Contrary to expectations, the Cosbo-3 mm-source identified by CARMA is
coincident with the NW-radio source and not the radio source (at
$z_\mathrm{photo}=2.4$)  closest to the mm-source identified
by MAMBO \citep[][]{bertoldi07}. Although consistent (within
$1\sigma$) with the MAMBO 1.1~mm flux, the CARMA 1.3~mm flux density
is somewhat lower. Thus, it may be possible that Cosbo-3 at $\sim11"$
resolution is itself a blend of two mm-sources, one of which was not detected
within the CARMA 1.3~mm sensitivity. Our CARMA observations put a
$3\sigma$ upper limit to the emission of a potential second mm-source
of $2.1$~mJy at 1.3~mm.

Cosbo-3 was found to be located in a strong overdensity ($30\times$
higher than the field) of star forming galaxies (Aravena et al.\
2010). Thus, it is possible that part of the MAMBO emission is distributed over several sources,  consistent with our CARMA observations. All the
galaxies in the overdensity have photometric redshifts in the range
$z=2.2-2.4$, providing strong statistical support to the photometric redshift of
our identified counterpart.

The mm-to-radio flux ratio suggests a redshift of
$z=3.2^{+0.6}_{-1.0}$. Our photometric redshift, based on the
deblended UV-MIR data, yields $z_\mathrm{phot}=1.9^{+0.9}_{-0..5}$,
which is consistent with the source's spectroscopic redshift (Riechers
et al., in prep.) and closer to the photometric redshift of the surrounding overdensity.  Assuming the UV-MIR based photometric-redshift value the
source's 20~cm radio flux density ($F_\mathrm{20cm}=78\pm13~\mu$Jy)
implies a radio luminosity of $\sim2\times10^{24}$~\wh \ and a SFR of
$\sim900$~\msolyr . Scaling an Arp~220 template we find an IR
luminosity of $L_\mathrm{IR}\sim1.5\times10^{13}$~\lsun . The redshift and
star formation rate of Cosbo-3 are fairly typical for SMGs, found to
form stars at similar rates and populating the redshift range $z=2-3$ (e.g.\
Chapman et al.\ 2005; Wardlow et al.\ 2010; Yun et al.\ 2011).

\subsubsection {Cosbo-8}

A source coincident with the position of Cosbo-8 is detected in the
radio band at high significance ($F_\mathrm{20cm}=104\pm13~\mu$Jy). As
in the case of the other two SMGs, in Spitzer images it is blended
with a source $1.2"$ to the NE (see \f{fig:stamps} \ and
\f{fig:deblendc3} ). Its deblended photometry is presented in
\t{tab:phot} \ and the deblended images in \f{fig:deblendc3} .

The mm-to-radio flux ratio suggests a redshift of
$z=1.9^{+0.5}_{-0.7}$. The UV-MIR-based photometric redshift is in the range of $z\sim3.6 - 4.3$ (using spectral models typical for AGN as this source is detected in the X-rays by Chandra). Assuming $z\sim4$ the radio
flux of Cosbo-8 implies a 20~cm luminosity of $\sim10^{25}$~\wh
\ and a radio-based SFR of $\sim6400$~\msolyr , an IR luminosity of
$L_\mathrm{IR}\sim1.1\times10^{13}$~\lsun \ (based on a scaled Arp~220
template), and an IR-based SFR of $\sim1700$~\msolyr \ (Bell 2003). The difference in the radio- and IR-based SFRs  suggests the
presence of an AGN also at radio wavelengths. 

Cosbo-8 corresponds to a point source detected at high significance in both the full C-COSMOS and the best-PSF C-COSMOS datasets with 15 (2.3 estimated background) and 5 (0 background) counts in the 0.5--2 keV band, respectively. Using the countrate-to-flux conversion factors from Puccetti et al.\ (2009), we obtain a flux of the source in the 0.5--2 keV band of $(3.8\pm1.1)\times 10^{-16}$ ergs s$^{-1}$ cm$^2$. Assuming  $z\sim4$ and $\Gamma=1.4$, this corresponds to a rest-frame luminosity in the 2--10 keV band of $6.8\times10^{43}$ ergs s$^{-1}$, which corresponds to the level of the emission of a typical AGN (using
$\Gamma=2$ results in a 10\% downward revision of the luminosity).

In the MAMBO-COSMOS area to-date only one other SMG (Cosbo-11) has been confirmed as a X-ray detected AGN (type-1 QSO at $z_\mathrm{spec}=1.83$; Aravena et al.\ 2008). Cosbo-11 is likely on-going a merger and shows radio and IR luminosities consistent with purely SF activity. Based on the duality of properties showing both properties of QSO and starbust it has been classified as a starburst-to-QSO 'transition' system. Cosbo-8, on the other hand, is also on-going a major starburst as implied by its IR luminosity, however, it shows an excess of radio emission with respect to that expected from the IR SED. This suggests that the AGN in this case is having a more important role in the bolometric output. Thus this source could also be classified as a starburst-QSO transition object, but possibly in a more advanced stage when the AGN starts to dominate the SED.

\section{Summary and Discussion}

In order to unambiguously determine the multi-wavelength counterparts
of three $\mathrm{F_{1mm}}>5.5$~mJy SMGs in the COSMOS field
(initially detected with MAMBO and AzTEC bolometers at low, $>10"$,
resolution), we performed interferometric observations at 1.3~mm and
$\sim2"-3"$ resolution using CARMA. The observations yielded $3-4\sigma$
detections coincident with positions of 20~cm radio sources
(VLA-COSMOS survey; \citealt{schinnerer07,schinnerer10}). Although all
three sources are coincident with radio detections, our observations
illustrate the need for high-resolution mm-imaging to determine
the correct counterparts of bolometer-identified SMGs. Without
high-resolution mm observations, the counterpart of Cosbo-3 would have
been misclassified as our observations associate this SMG with the
radio source (out of two radio-sources) within the MAMBO beam
that is further away from the MAMBO source center.

All three sources identified here are blended in the MIR by nearby
bright galaxies. We have carefully deblended their photometry, and
derived photometric redshifts.  We find photometric redshift of $z_\mathrm{phot}\mathrm{(AzTEC/C1)}=5.6\pm1.2$, $z_\mathrm{phot}\mathrm{(Cosbo-3)}=1.9^{+0.9}_{-0.5}$, and $z_\mathrm{phot}\mathrm{(Cosbo-8)}\sim4$. These are consistent with
mm-to-radio-flux based estimates for AzTEC/C1 and Cosbo-3 ($4.3^{+0.7}_{-1.4}$ and $3.2^{+0.6}_{-1.0}$, respectively), but inconsistent with that inferred for Cosbo-8 ($1.9^{+0.5}_{-0.7}$). This is naturally understood as (part of) the radio flux in Cosbo-8 may arise from the associated AGN identified via the X-ray Chandra detection. An increased radio flux due to processes not related to star formation would lead to an artificial decrease in the mm-to-radio-flux based redshift. 

Our three SMGs seem to show a relatively large redshift
spread, comparable to optically selected SMGs, but with a potential
bias towards higher redshifts as all 3 SMGs have been found to be at
$z\gtrsim2$ (e.g.\ Chapman et al.\ 2005). In general, although it has
been shown that the SMG population peaks between redshifts 2 and 3
(e.g.\ Chapman et al.\ 2005; Wardlow et al.\ 2011; Yun et al.\ 2011),
their exact redshift distribution (and thus their cosmic evolution) is
still rather poorly understood. This is mainly related to statistical
counterpart selection biases induced by the large single-dish mm-beams
(see e.g.\ Yun et al.\ 2011 for a more detailed discussion). This can
be avoided by mm-interferometric imaging at intermediate/high
angular resolution of {\em complete} samples of SMGs. However, generating
such samples has been a very time-consuming, and largely unfeasible
process, and assembling complete samples of SMG counterparts (and
their redshifts) will require surveys with facilities such as ALMA and
LMT. Nonetheless, existing mm-interferometric observations of SMGs
already suggest that a fraction of these sources (at least at the
bright end) is unexpectedly at redshifts $z\gtrsim4$ (e.g.\ 4/17
AzTEC/JCMT SMGs detected by SMA/VLA are spectroscopically confirmed to
be at $z\gtrsim4$; Scott et al.\ 2008; Younger et al 2007, 2009; Capak
et al.\ 2008, Schinnerer et al.\ 2008; Riechers et al.\ 2010; Capak et
al.\ 2010; Karim et al., in prep). In this cosmic epoch to-date there
are only about ten SMGs confirmed
\citep{daddi09a,daddi09b,capak08,schinnerer08,riechers10,capak10,
  smo11,coppin09,coppin10,knudsen10,cox11,combes12}. 
  Given their large star formation
rates, these very high redshift SMGs are considered to be optimal
candidates for the progenitors of $z\sim2$ massive red galaxies
(Cimatti et al.\ 2008). However, as this population is just starting to emerge their
role in galaxy evolution is still largely unexplored, and efforts to
identify such sources and characterize their properties is critical. Based on
mm-/radio-interferometry we have associated AzTEC/C1, the brightest
SMG in the AzTEC/ASTE COSMOS survey, with a MIR/radio source at
$z\gtrsim4$. Interestingly, its high-redshift, mm- and
radio-fluxes are comparable to that of AzTEC-1 -- the brightest SMG in
the AzTEC/JCMT COSMOS survey (Scott et al.\ 2008) -- with properties
resembling those expected for the progenitors of compact massive red
galaxies at $z\sim2$ (\smo\ et al.\ 2011).

Studies of SMGs, and the formation of passive red galaxies (often
found in the most massive galaxy clusters) suggest that the first may
be progenitors of the second (e.g.\ Michalowski et al.\ 2010; Hickox
et al.\ 2011).  Consistent with this picture, in which strong
clustering of SMGs is expected, spatial clustering analysis of
$z\lesssim3$ SMGs find that SMGs cluster strongly (e.g.\ Hickox et al.\ 2011), and that they can
be statistically associated with galaxy overdensities (Aravena et al.\
2010). Furthermore, only recently have two $z>4$ protoclusters
hosting SMGs been identified (Daddi et al.\ 2009; Capak et al.\ 2010),
providing valuable laboratories to study structure formation at the
earliest cosmic times. Based on $BzK$-selected galaxies Aravena et
al.\ (2010) identified a significant galaxy overdensity at $z\sim2.3$
in the area around Cosbo-3. The mm/radio-interferometric observations
and photometric redshift computation, presented here, strengthen the
case that Cosbo-3 is indeed associated with this galaxy overdensity (rather
than being a fore-/back-ground galaxy). This system therefore proves
interesting for further studies of the dense environment of SMGs at the
peak epoch of this population, linking the properties of $z>4$
proto-clusters hosting SMGs and local galaxy clusters.

In summary, we have identified the counterparts of three single-dish
detected SMGs in the COSMOS field via mm-interferometry, and presented
their (deblended) UV-MIR photometry, and redshift estimates. Such
studies are an important step towards reaching unbiased statistical
samples of SMGs with accurately determined counterparts and redshifts
-- a necessary but yet unaccomplished prerequisite for comprehensive
studies of physical properties of SMG population(s) and their role
in galaxy formation and evolution.

\acknowledgments The authors than the anonymous referee for comments that significantly improved the manuscript. The research leading to these results has received
funding from the European Union's Seventh Framework programme under
grant agreement 229517. The National Radio Astronomy Observatory is a facility of the National Science Foundation operated under cooperative agreement by Associated Universities, Inc. MS acknowledges support by the German Deutsche Forschungsgemeinschaft, DFG Leibniz Prize (FKZ HA 1850/28-1). AK acknowledges support by the DFG grant SCHI 536/3-3 as part of the Priority Programme 1177 (Witnesses of Cosmic History: Formation and Evolution of Black Holes, Galaxies and their Environment). Support for CARMA construction was derived
from the Gordon and Betty Moore Foundation, the Kenneth T. and Eileen
L. Norris Foundation, the James S. McDonnell Foundation, the
Associates of the California Institute of Technology, the University
of Chicago, the states of California, Illinois, and Maryland, and the
National Science Foundation. Ongoing CARMA development and operations
are supported by the National Science Foundation under a cooperative
agreement, and by the CARMA partner universities. 

{}

\end{document}